\documentclass[%
 aip,
 amsmath,amssymb,
 reprint,%
]{revtex4-1}
\usepackage{graphicx}
\usepackage{dcolumn}
\usepackage{bm}
\usepackage{tikz}
\usetikzlibrary{angles}
\usetikzlibrary{quotes}
\usepackage[utf8]{inputenc}
\usepackage[T1]{fontenc}
\usepackage{mathptmx}
\usepackage{amsmath,amssymb}

\begin{document}

\preprint{AIP/123-QED}
\title{From period to quasi-period to chaos: A continuous spectrum of orbits of charged particles trapped in a dipole magnetic field} 



\author{Yuxin Xie}
\altaffiliation{Also at Phillips Academy Andover, 180 Main Street, Andover, MA, 01810, USA}
\author{Siming Liu}\email{liusm@pmo.ac.cn}
\affiliation{Key laboratory of dark matter and space astronomy, Purple Mountain Observatory, Chinese Academy of Sciences, Nanjing 210023, People's Republic of China}





\date{\today}

\begin{abstract}
Via evaluation of the Lyapunov exponent, we report the discovery of three prominent sets of phase space regimes of quasi-periodic orbits of charged particles trapped in a dipole magnetic field. Besides the low energy regime that has been studied extensively and covers more than $\sim 10\%$ in each dimension of the phase space of trapped orbits, there are two sets of high energy regimes, the largest of which covers more than $\sim 4\%$ in each dimension of the phase space of trapped orbits. Particles in these high energy orbits may be observed in space and be realized in plasma experiments on the Earth.
\end{abstract}

\maketitle 

\begin{quotation}
It is well-known that there are quasi-periodic orbits around stable periodic orbits in Hamiltonian systems with 2 degrees of freedom \citep{1990ComPh...4..549S} and these quasi-periodic orbits are stable as well \citep{1963RuMaS..18....9A}. Since periodic orbits appear to have a negligible measure in the phase space \citep{1975Ap&SS..32..115M,1978CeMec..17..215M, 1990CeMDA..49..327J}, they are difficult to realize in nature. Quasi-periodic orbits, on the other hand, may occupy a finite volume in the 4 dimensional (4D) phase space and be readily detectable.
A chaotic orbit has at least one positive Lyapunov exponent. The Lyapunov exponents of quasi-periodic orbits, on the other hand, should be zero \citep{doi:10.1142/S0218127400000177}. Via calculation of the Lyapunov exponent of orbits of trapped charged particles in a dipole magnetic field, we scanned the corresponding phase space and found several prominent regimes of quasi-periodic orbits associated with stable periodic orbits in the equatorial plane \citep{devogelaere1950}. {\bf These regimes appear to be connected to some small regimes of quasi-periodic orbits associated with stable periodic orbits in the Meridian plane.}
Our numerical results also show a continuous spectrum of these orbits from stable periodic, to quasi-periodic with vanishing Lyapunov exponents, and eventually to chaotic ones with at least one positive Lyapunov exponent and there are unstable periodic orbits with a positive maximum Lyapunov exponent.
\end{quotation}

\section{\label{sec:level1} Introduction 
}

Motions of a charged particle in a dipole magnetic field were first systematically investigated by \citet{1930ZA......1..237S}. Due to the axis symmetry of the system, it can be reduced to a 2D Hamiltonian system and is later called the St\"{o}rmer problem for its broad implications in mathematics, space science, and physics \citep{1965RvGSP...3..255D,1990ComPh...4..549S, doi:10.1063/1.859308, 2016PhRvE..94d3203S, NaokiKENMOCHI2019}. To better understand the Van Allen radiation belt discovered in the mid-20th century, there have been renewed interests in solving the St\"{o}rmer's problem. In particular, periodic orbits in the equatorial plane have been explored extensively \citep{devogelaere1950}, and there are also a few branches of periodic orbits in the Meridian plane at relatively high energies with distinct orbital shapes  \citep{1975Ap&SS..32..115M,1978CeMec..17..215M,1990CeMDA..49..327J}. At low energies, due to the presence of approximate adiabatic invariant, the problem can be further simplified \citep{1963RvGSP...1..283N, doi:10.1029/JA081i013p02327}, and the results, the so-called guiding center approximation, have been applied to studies of pulsars, radiation belts, and dynamics of particles in magnetic confinement devices \citep{2018PhRvL.121w5003H, 2014Natur.515..531B, NaokiKENMOCHI2019}.

It is well-known that particles moving in a dipole magnetic field are either trapped or can escape to infinity \citep{doi:10.1142/S0218127400000177}. For those trapped particles, in which people are mostly interested, their motions can be divided into three categories: quasi-periodic, chaotic, or hyper-chaotic depending on the number of positive Lyapunov exponents they have \citep{doi:10.1142/S0218127400000177}. Stable periodic orbits can be considered as special cases of quasi-periodic orbits. It has been suggested that the type of motion of trapped particles depends primarily on its energy. Particles with low energies will have quasi-periodic orbits, those with intermediate energy will move in a chaotic manner with a positive Lyapunov exponent, while high energy particles will display hyper-chaotic motion with two positive Lyapunov exponents.

The main goal of this paper is to have a quantitative classification of orbits of trapped particles with different initial conditions in the 4D phase space via evaluation of their Lyapunov exponents. This study also clarifies the relations among stable periodic orbits,  quasi-periodic orbits, and chaotic orbits, revealing a spectrum of orbits of trapped particles evolving continuously with the initial conditions.

\section{The St\"{o}rmer Problem}
$\phi$

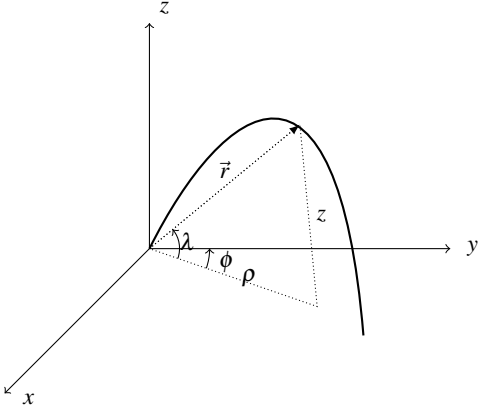
\begin{figure}[htb]
    \centering
\begin{tikzpicture}
\draw[->] (-2.5,0) (0,0) -- (4,0) node at (4.3,0,0) {$y$};
\draw[->] (0,0) -- (0,3,0) node at (0.2,3.2,0) {$z$};
\draw[->] (0,0) -- (0,0,5) node at (0.4,0,5.2) {$x$};
\draw[densely dotted, -latex] (0,0,0) -- (2,1.65,0);
\draw [densely dotted] (0,0,0) -- node[above] {$\vec{r}$} (2,1.65,0)  -- node[right] {$z$}(3,0,2)--node[right] {$\rho$}cycle ;
\draw [thick](0,0,0)..controls (1,2,0)and(2.5,3,0) ..(4,0,3);
\draw (3,0,2) coordinate (A)
 (0,0) coordinate (B)  (2,0,0)coordinate (C)
pic [draw,->, "$\phi$", angle eccentricity=1.3,angle radius= 0.8cm] {angle};
\draw (3,0,2) coordinate (A)
(0,0) coordinate (B)  (2,1.65,0)coordinate (C)
pic [draw,->, "$\lambda$", angle eccentricity=1.3,angle radius= 0.4cm] {angle};
\end{tikzpicture}
\caption{Coordinates adopted in this paper. The solid line indicates a magnetic field line and the dashed lines indicate the vector ${\bf r}$ and its components.}
    \label{fig:coordinates}
\end{figure}

For the sake of completeness, we briefly summarize the St\"{o}rmer problem here.
We will adopt the coordinates as shown in Figure \ref{fig:coordinates} with a magnetic dipole at the origin pointing to the $z$ direction. {\bf The position angles of the vector $\vec{r}$ are given by $\lambda$, and $\phi$. Its projection onto the \{$x, \ y$\} plane has a length of $\rho=r\cos{\lambda}$.}
Then the vector magnetic potential $\vec{A}$ can be put as the following:
  \begin{align}
  \vec{A}(\vec{r})&=\frac{\mu_0}{4\pi} \frac{\vec{\mu} \times \hat{r}}{r^2}  \ \ \ \ \  {\rm with} \\[1ex]
  \vec{\mu} &= \left(\begin{array}{c}
                 0,
                 0,
                 M
               \end{array}\right)  \ \ \ \ \ {\rm and} \\[1ex]
  \hat{r} &= \left(\begin{array}{c}
                 \cos \lambda \cdot \sin \phi,
                 \cos \lambda \cdot \cos \phi,
                 \sin \lambda
               \end{array}\right) 
    \end{align}
where $\mu_0$ is the magnetic permittivity of the vacuum and $M$ is the magnetic moment of the dipole. Then we have
\begin{equation}
    \vec{A}  = \frac{\mu_0}{4\pi} M \frac{\rho}{r^3} \hat{\phi}.
\end{equation}
and magnetic field lines are given by $r=r_0 \cos^2 \lambda$, where $r_0$ specifies the lines and $r^2=\rho^2+z^2$

For a particle with a speed of $v$, the Lorentz factor is given by
\begin{equation}
    \gamma=\left[ 1-\left(\frac{v^2}{c^2}\right) \right]^{-\frac{1}{2}}
\end{equation}
where $c$ is the speed of light. Then
the relativistic Hamiltonian of a charged particle with a charge $q$ moving in the above dipole magnetic field is given by:
\begin{equation}
    H_0=\left\{m_0^2c^4+c^2\left[P_z^2+P_\rho^2+\left(\frac{P_\phi}{\rho}-qA_\phi\right)^2\right] \right\}^{\frac{1}{2}}
\end{equation}
where $m_0$ is the rest mass of the charged particle, and
$P_z =\gamma m_0\dot{z}$, $P_\rho =\gamma m_0 \dot{\rho}$, $P_\phi = \gamma m_0 \rho^2 \dot{\phi}+q\rho A_\phi$, where an upper dot indicates derivative with respect to time $t$.
Since the magnetic force does not change the particle energy, the particle's speed remains constant. The canonical angular momentum along the $z$-axis $P_\phi$ is also a constant for the independence of $H$ on $\phi$.

The problem can be simplified by adopting the following Hamiltonian
\begin{equation}
    H=\frac{1}{2\gamma m_0}\left[P_z^2+P_\rho^2+\left(\frac{P_\phi}{\rho}-qA_\phi\right)^2\right].
\end{equation}
Considering the energy conservation and independence of $H$ on $\phi$, the Hamiltonian can be reduced to
\begin{equation}
    H=\frac{1}{2m}(P_z^2+P_\rho^2) +V
\end{equation}
with the effective potential
\begin{equation}
    V=\frac{1}{2m}\left(\frac{P_\phi}{\rho}-qA_\phi\right)^2\,,
\end{equation}
where $m =\gamma m_0$.

A characteristic length scale $L$ can be introduced with $P_\phi$: $L = q\mu_0 M/4\pi P_\phi$.
The corresponding energy scale is given by $P_\phi^2/mL^2\propto L^{-4}$, and the characteristic momentum is given by $P_\phi/L$.
Then we get the following dimensionless Hamiltonian:
\begin{eqnarray}
    h&=&\frac{1}{2}\left[{p_z}^2+{p_\rho}^2+\left(\frac{1}{\rho}-\frac{\rho}{r\,^3}\right)^2\right] \nonumber \\
    &=& \frac{1}{2}\left({p_z}^2+{p_\rho}^2\right)+ V ~~
\end{eqnarray}
with the effective potential
\begin{equation}
    V=\frac{1}{2}\left(\frac{1}{\rho}-\frac{\rho}{r\,^3}\right)^2.
\end{equation}
And the Hamiltonian's equations are: 
       \begin{align}
         \dot{z}&= \frac{\partial h}{\partial p_z}=p_z\,, \\[1ex]
  \dot{\rho}&=\frac{\partial h}{\partial p_\rho}=p_\rho\,, \\[1ex]
  \dot{p_z}&= -\frac{\partial h}{\partial z}=-\frac{3z\rho\left[\frac{1}{\rho}-\frac{\rho}{(z^2+\rho^2)^\frac{3}{2}}\right]}{(z^2+\rho^2)^\frac{5}{2}}\,,  \\
      \dot{p_\rho} &=-\frac{\partial h}{\partial \rho} \\[1ex] \nonumber
  &=\left[ \frac{1}{\rho^2}-\frac{3\rho^2}{(z^2+\rho^2)^\frac{3}{2}}
  +\frac{1}{(z^2+\rho^2)^\frac{3}{2}}\right]\left[\frac{1}{\rho}-\frac{\rho}{(z^2+\rho^2)^\frac{3}{2}}\right].
  \end{align} 
For given initial conditions, these equations can be solved numerically to obtain the particle orbit in the phase space.

Orbits of particles with chaotic motions are usually very sensitive to the initial conditions. After a long period of motion starting with the initial conditions, the orbit of motion deviates greatly from the initial path that it began with. Lyapunov exponent, which characterizes the separation rate between infinitely close trajectories, helps us determine to what extent the resulting trajectory depends on changes in the initial conditions. {\bf For an n-D dynamical system with $\dot{x}_i = f_i(x)$, there are n Lyapunov exponents, which are assigned to each of the free variables. The Lyapunov exponents are defined as the eigenvalues of the following matrix $\Lambda$:
\begin{equation}
    \Lambda = \lim_{t\rightarrow\infty}{1\over 2t}\ln [Y(t)Y^T(t)]
\end{equation}
where $\dot{Y}(t) = J(t)Y(t)$, $Y(0)$ is the unit matrix, and $J(t)$ is the Jacobian matrix of the system:
\begin{equation}
    J_{ij}(t) = {\partial f_i[x(t)]\over \partial x_j}\,.
\end{equation}
}

In our case, with \{$z$, $\rho$, $p_z$, $p_\rho$\}, there are four Lyapunov exponents in total. For a Hamiltonian system, the sum of the Lyapunov exponents of all dimensions in the phase space must be zero and the Lyapunov exponents must be in pairs of opposite sign. Therefore there are at most two positive Lyapunov exponents.
A positive maximum Lyapunov exponent is often considered as a definition of deterministic chaos. If the maximum Lyapunov exponent is not greater than 0, then the particle has a quasi-periodic orbit. If there exists two positive Lyapunov exponents, then the particle's orbit is hyper-chaotic \citep{doi:10.1142/S0218127400000177}. For the time series of coordinates of a Hamiltonian system obtained numerically above, the corresponding Lyapunov exponents can be readily obtained \citep{1980Mecc...15....9B, 1985PhyD...16..285W}. {\bf Computations of the Lyapunov exponents have been performed in Matlab with integrator ode113 using absolute tolerance of
$10^{-10}$ and a relative tolerance of $10^{-6}$. Jacobians needed for Lyapunov exponent calculations are found analytically. The Poincar\'{e} maps are obtained with Mathematica with an absolute tolerance of $10^{-10}$ and a relative tolerance of $10^{-8}$ unless specified otherwise.}

\section{Results}

\begin{figure*}[htbp]
\includegraphics[width=1.1\columnwidth, angle=0.0]{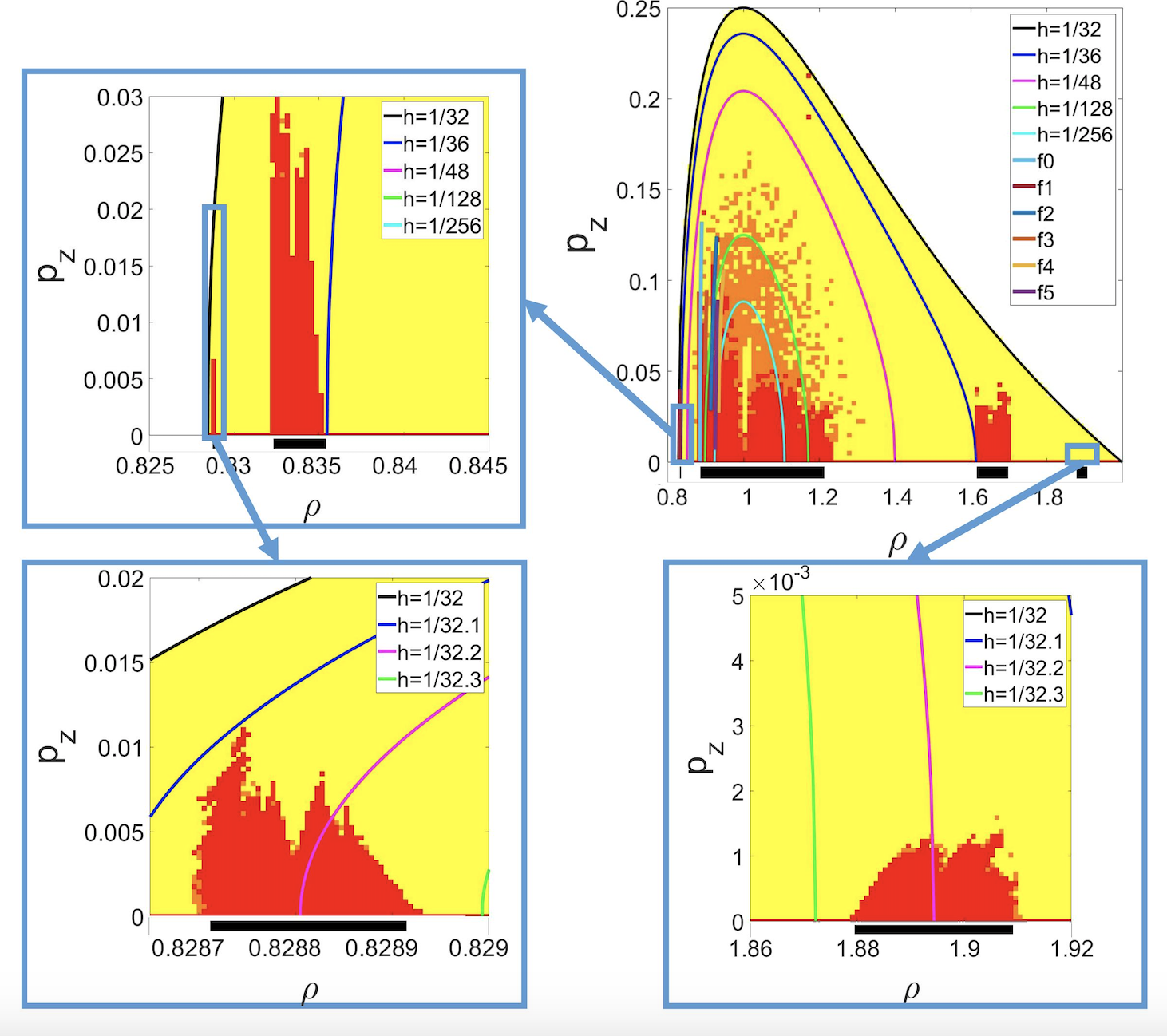}\\
\caption{\label{fig1a} Distribution of the number of positive Lyapunov exponents in the phase space of \{0, $\rho$, $p_z$, 0\}. Red, orange, and yellow corresponds to 0, 1, and 2, respectively. Colored curves show the energy (indicated by h) contours as indicated in these figures. The line segments labeled as f0-f5 correspond to stable periodic orbits \citep{1978CeMec..17..215M}. The thick horizontal bars below the $\rho$ axis indicate stable periodic orbits in the equatorial plane \citep{devogelaere1950}. The upper-left and lower-right panels are enlargement of two rectangular regions in the upper-right panel. The lower-left panel is enlargement of a rectangular region in the upper-left panel. The first set of high-energy quasi-periodic orbits locates just above h $=1/36$ (blue line in the upper panels). The second set locates around h $=1/32.2$ (pink lines in the lower panels). See text for details.}
\end{figure*}

\begin{figure*}[htbp]

\includegraphics[width=1.05\columnwidth, angle=0.0]{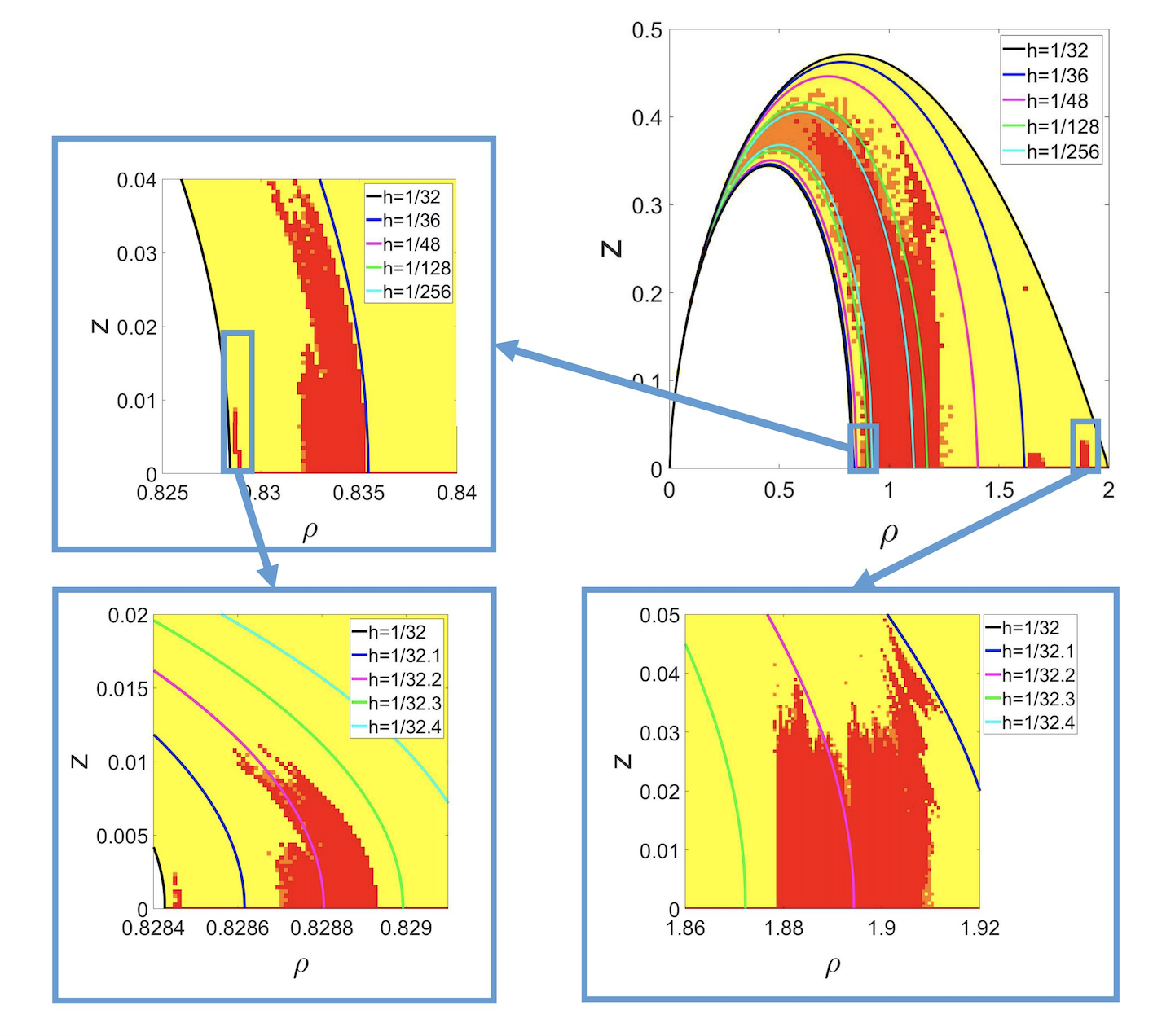}\\

\caption{\label{fig1b} The same as Figure \ref{fig1a} but for the phase space of \{$z$, $\rho$, 0, 0\}. A third set of quasi-periodic orbits can be seen in the lower left panel below h $=1/32$.
}
\end{figure*}

\begin{figure*}[!htbp]

\includegraphics[width=0.858\columnwidth, angle=0.0]{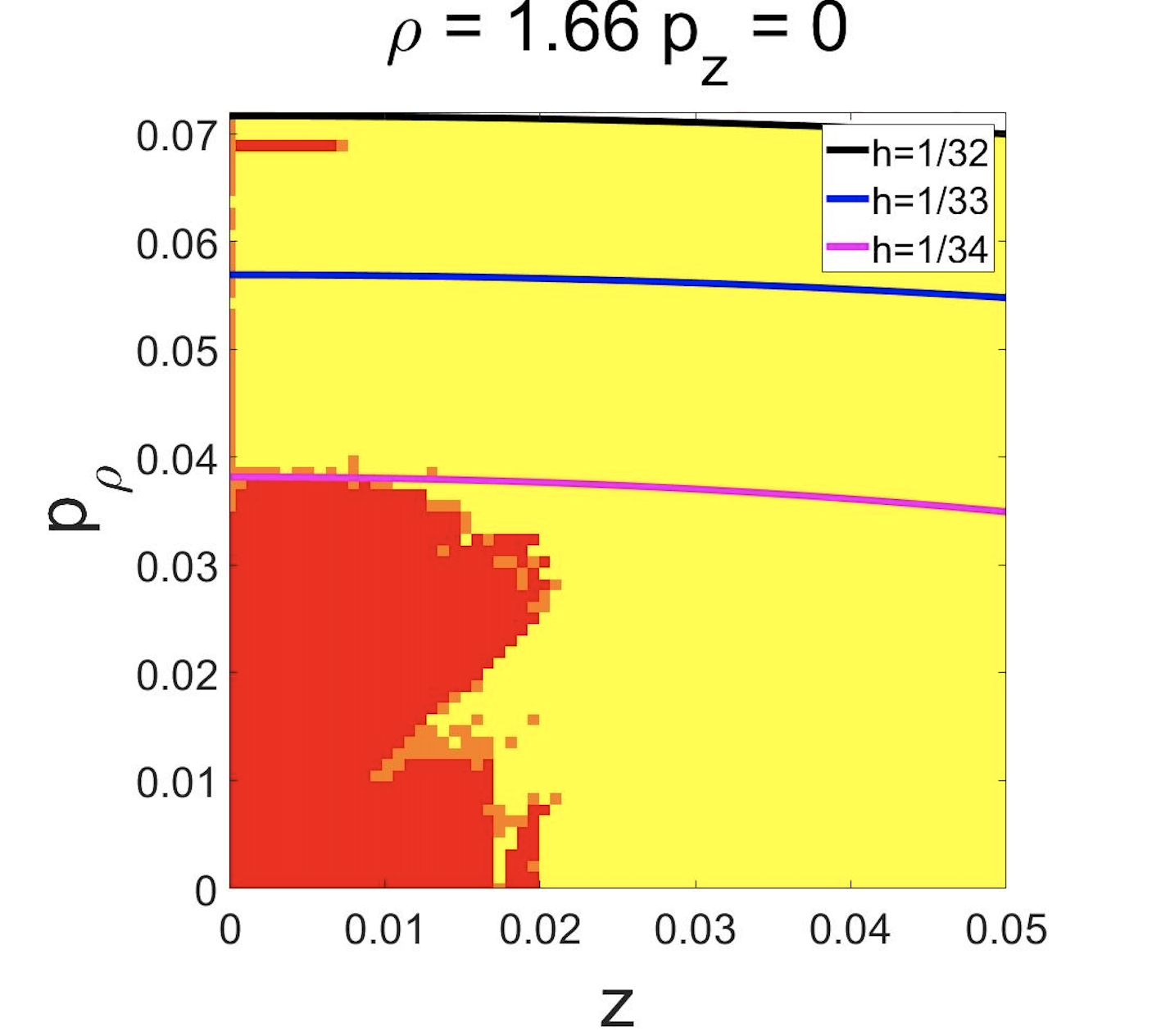}
\includegraphics[width=0.82\columnwidth, angle=0.0]{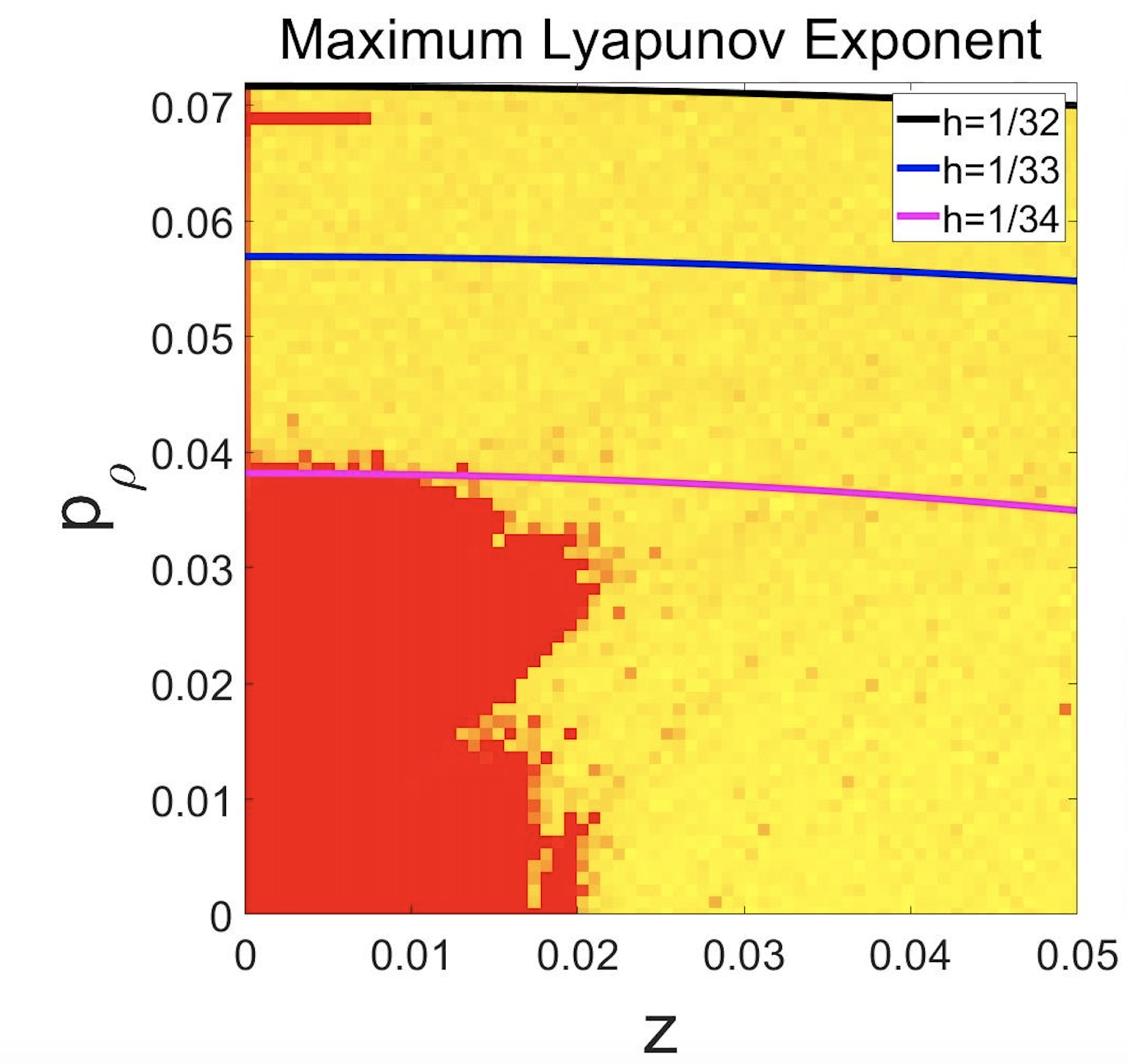} 
\caption{\label{fig1c} Left: the same as Figure \ref{fig1a} but for the phase space of \{$z$, 1.66, 0, $p_\rho$\} through the first high-energy set. The right panel shows the distribution of the maximum Lyapunov exponent of a rectangular region in the left panel. }
\end{figure*}

\begin{figure*}[!htbp]
\includegraphics[width=1.0\columnwidth, angle=0.0]{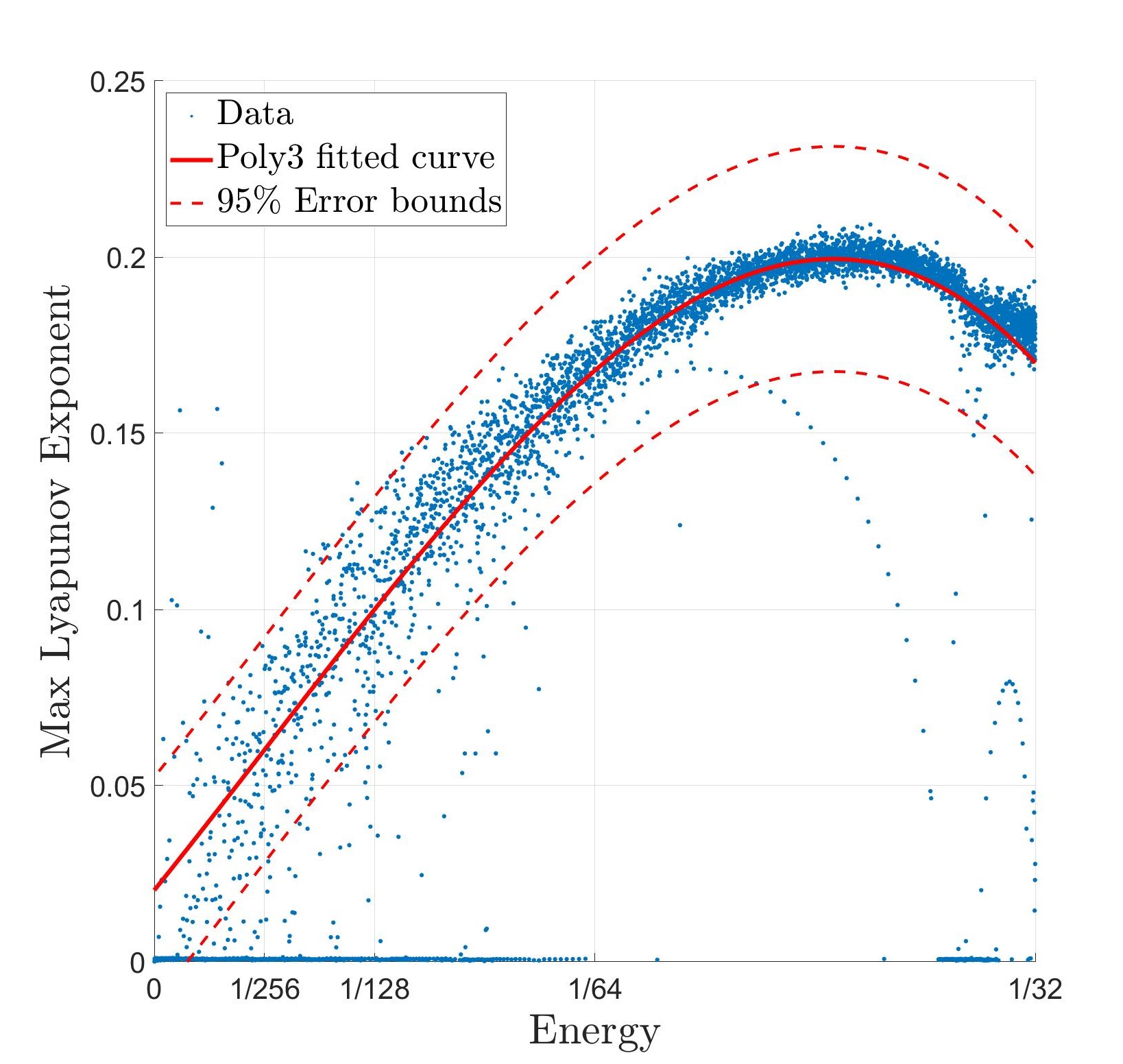}

\caption{\label{fig01} Correlation between the maximum Lyapunov exponent and energy h in the phase space of \{0, $\rho$, $p_z$, 0\} for the initial conditions. The dotted curves below the main distribution correspond to unstable periodic orbits in the equatorial plane. The red lines show a fit to the correlation with polynomial function of degree 3 and the error bands.}
\end{figure*}

\begin{figure*}[htbp]
\includegraphics[width=0.43\linewidth, angle=0.0]{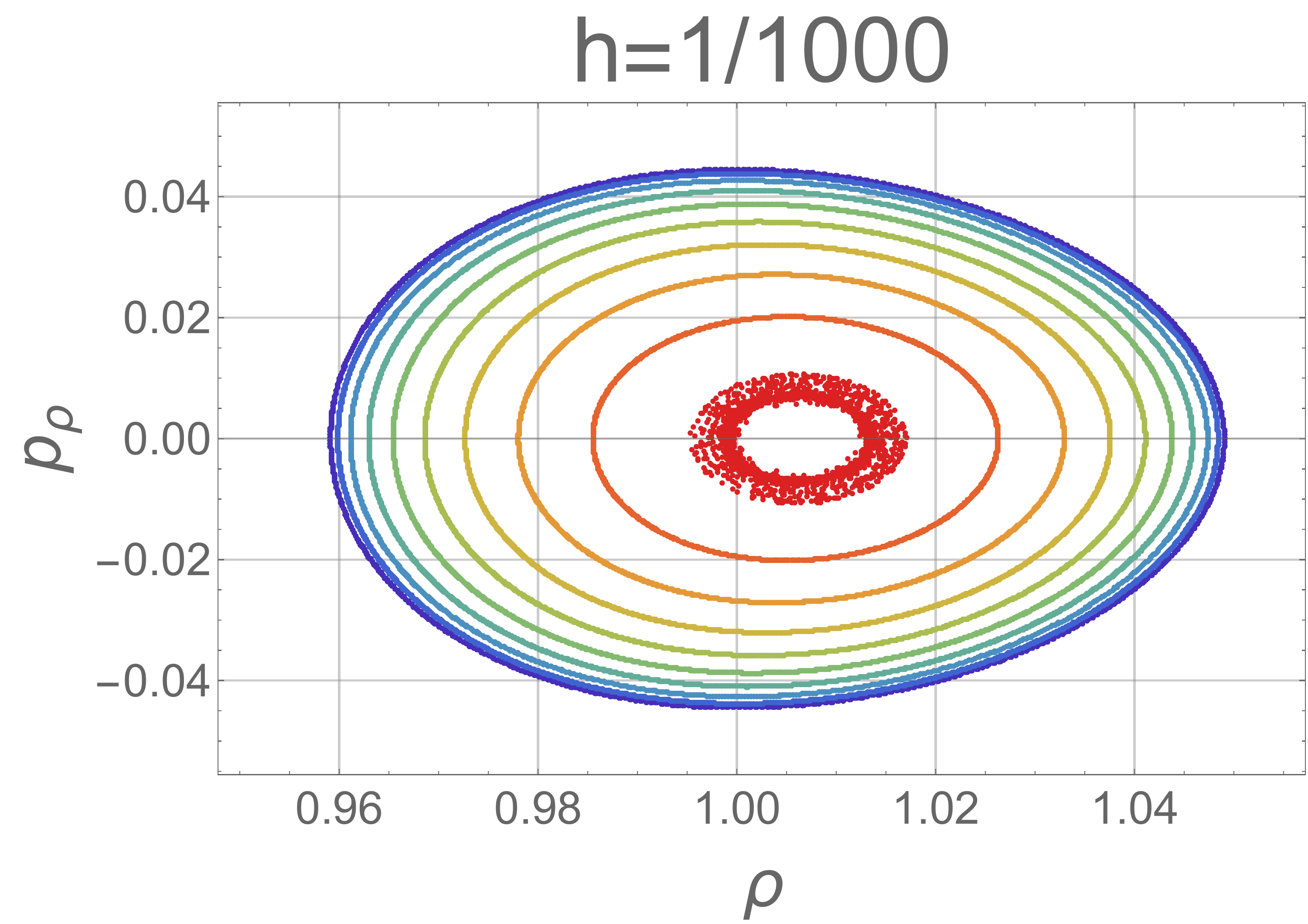}
\includegraphics[width=0.43\linewidth, angle=0.0]{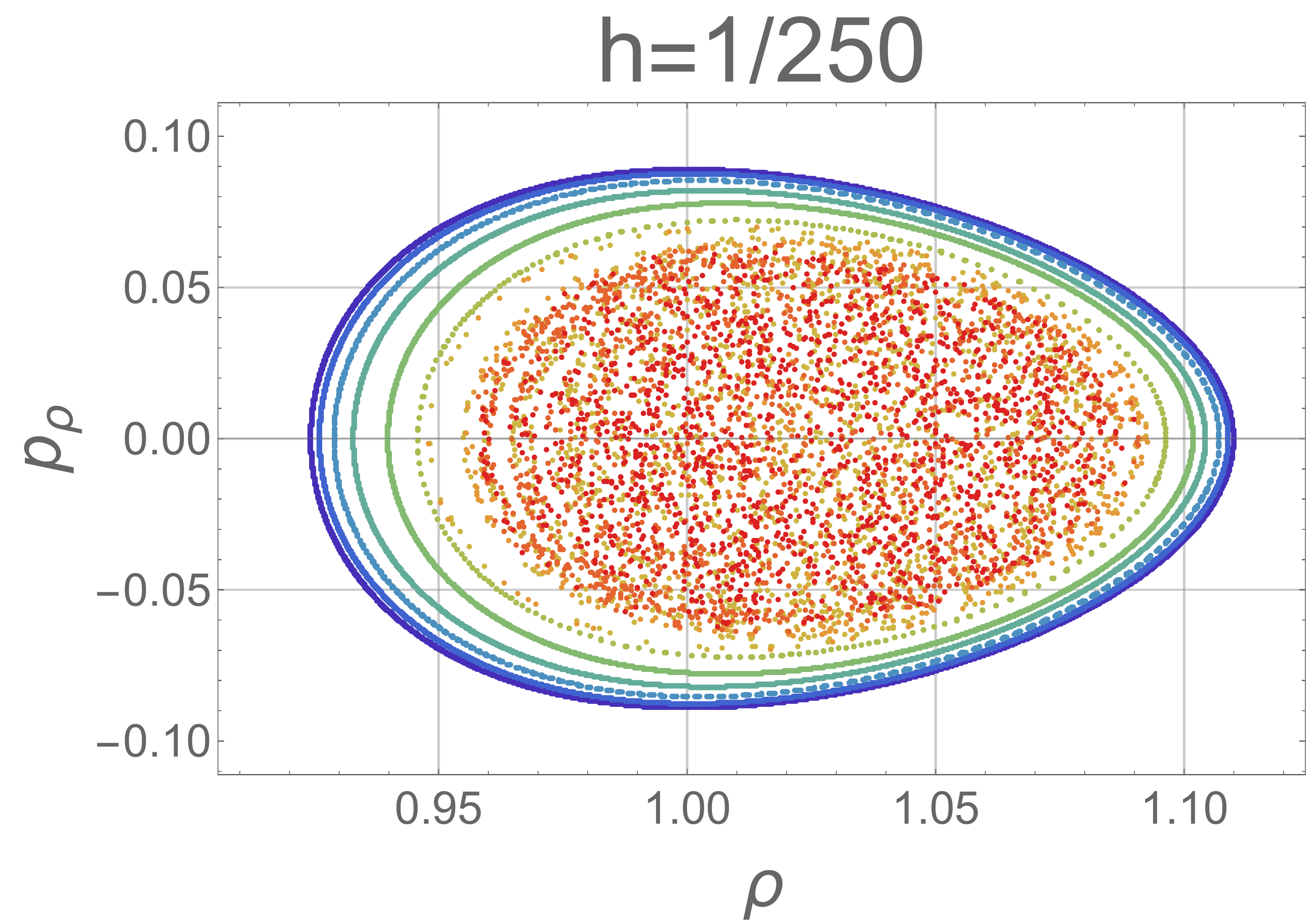}\\
\includegraphics[width=0.43\linewidth, angle=0.0]{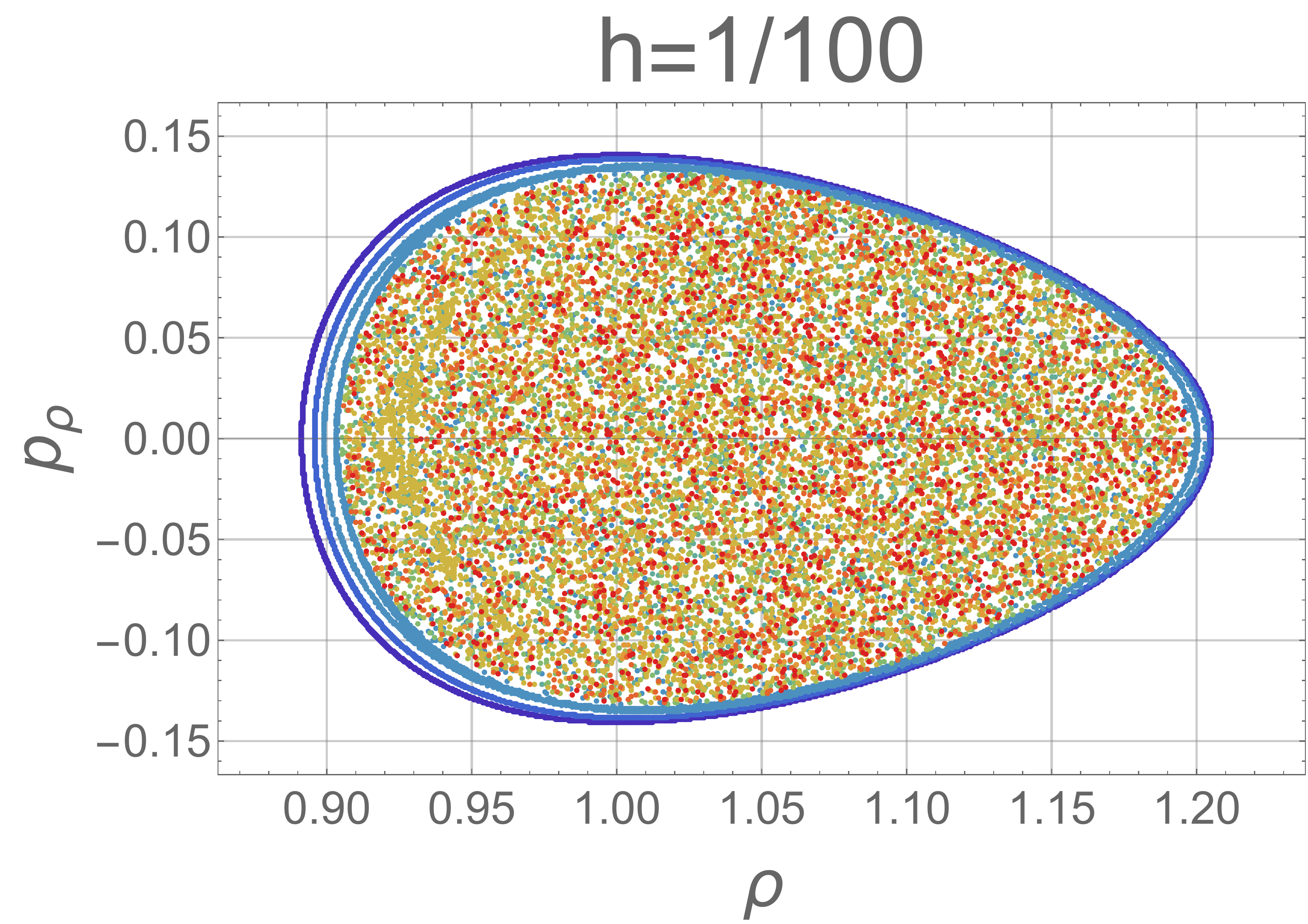}
\includegraphics[width=0.43\linewidth,angle=0.0]{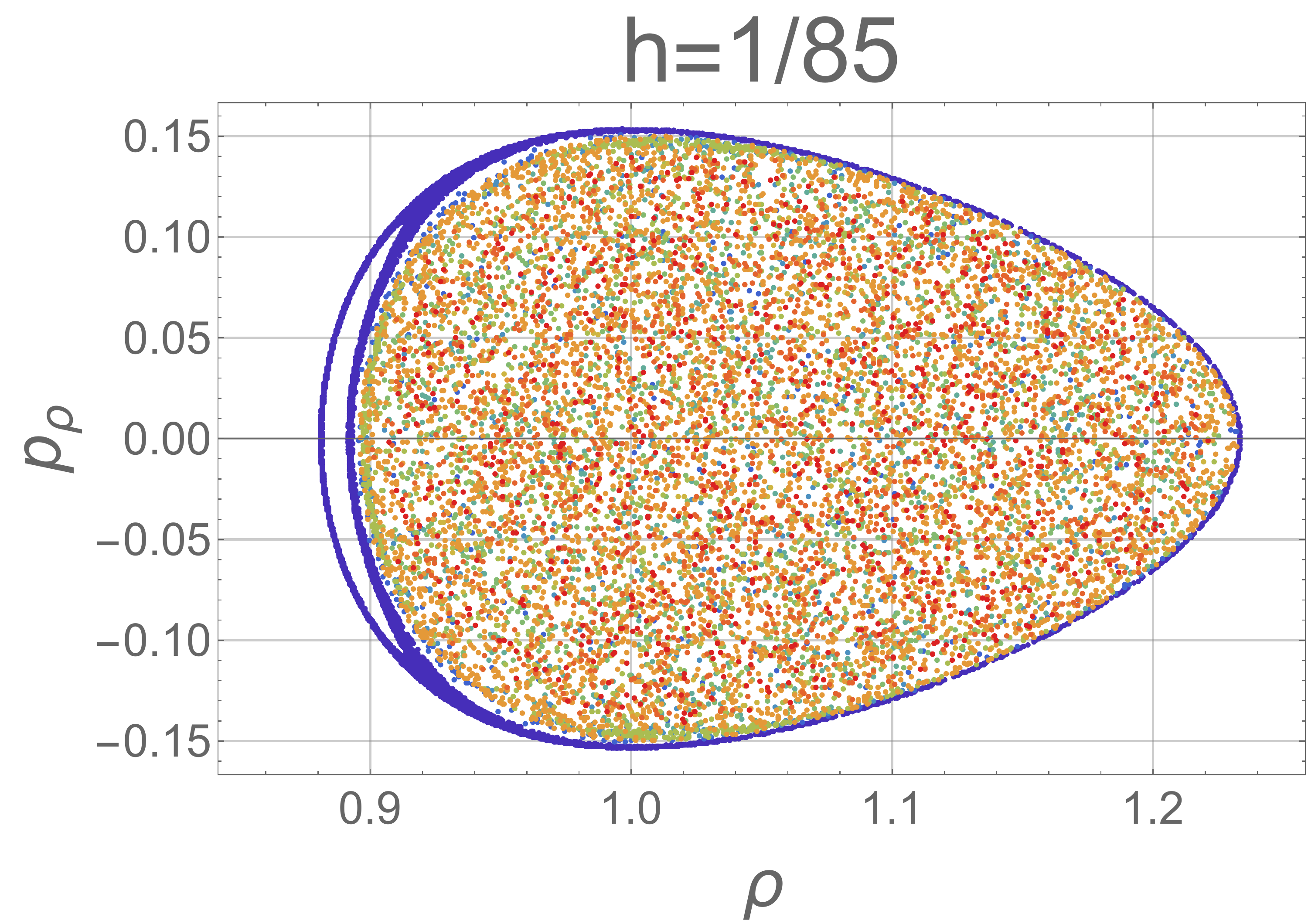}\\

\caption{\label{fig12} Poincar\'{e} maps of particles injected from the $z=0$, and $\rho=1$ phase plane with the energy indicated on the top of each panel (See also Fig. \ref{fig1a}). We show ten particles with different initial pitch angles separated by 0.1 in their sine starting from 0. These particles are indicated with different colors. Orbits in the equatorial plane do not have a Poincar\'{e} map.}
\end{figure*}

{\bf It is well-known that the energy level of h$=1/32$ separates the orbits into trapped and scattering regimes \citep{doi:10.1063/1.859308}. Although there are trapped periodic orbits beyond the $1/32$ energy level \citep{1978CeMec..17..215M}, they have a zero measure. Our numerical results show all other orbits beyond the $1/32$ energy level are scattering orbits that extend to infinity. We therefore focus on studying these trapped orbits with h$\leq1/32$.}

The key results of this paper are shown in Figures \ref{fig1a}-\ref{fig1c}, where we scan planes of initial conditions with $z=0$ and $p_\rho=0$ (Fig. \ref{fig1a}), $p_\rho=0$ and $p_z=0$ (Fig. \ref{fig1b}), and locally with $\rho = 1.66$ and $p_z=0$ (Fig. \ref{fig1c}). Here the red color indicates regions with vanishing Lyapunov exponents, corresponding to quasi-periodic orbits. The orange and yellow colors represent regions with one and two positive Lyapunov exponents, respectively. {\bf Fractal structure can be seen at the boundaries of regions for quasi-periodic orbits, which is typical for dynamical systems.}

We also indicate contours of different energy levels represented by the values of h. It can be seen that in general, orbits with low energies tend to have vanishing Lyapunov exponents corresponding to quasi-periodic orbits. With the increase of energy h, the orbits become more chaotic as pointed out by \citet{doi:10.1142/S0218127400000177}. {\bf However, Figure \ref{fig01} shows that there is no simple correlation between the maximum Lyapunov exponent and the energy, which is compatible with the right panel of Figure \ref{fig1c}, where one can see that between different energy contours, the distribution of the maximum Lyapunov exponent can spread over a broad range.}

Within the 1/32 energy contour, we discovered two sets of high-energy regions with vanishing Lyapunov exponents (Figs. \ref{fig1a}-\ref{fig1c}). These sets are connected via the energy contours indicating that they may represent the same orbits. For the convenience of discussion, we will rank these sets according to the energy with the lower energy one right above h $=1/36$ ranking the first. The structure of these regions in three independent planes of the initial conditions shows that these regions are 4D structures (Figs. \ref{fig1a}-\ref{fig1c}). The largest region of the first set covers more than 4\% in each dimension within the 1/32 energy contour. The second set locates around h $=1/32.2$ near the equatorial plane. 

{\bf In Figure \ref{fig1a}, we use thick black line segments to indicate stable periodic orbits in the equatorial plane and solid line segments to indicate stable periodic orbits in the Meridian plane that belong to families of periodic orbits f0-f5 introduced by \citet{1978CeMec..17..215M}. It can be seen that quasi-periodic orbits appear to be always associated with stable periodic orbits. 
However, quasi-periodic orbits associated with stable periodic orbits in the Meridian plane have a much smaller volume.  For $h>1/32$, we found that only periodic orbits are bounded. Particles in other orbits will escape to infinity no matter what values of the maximum Lyapunov exponent they have.}
Near unstable periodic orbits, there is no quasi-periodic orbit with vanishing Lyapunov exponents. Figure \ref{fig01} shows that the maximum Lyapunov exponents of these unstable periodic orbits actually follow some pattern (dotted curves), rising first then declining with the increase of energy h between regions of stable periodic orbits. {\bf Therefore although periodic orbits are relatively simple, they may have a positive Lyapunov exponent. A positive Lyapunov exponent is a necessary but not sufficient condition for chaotic trajectories.}

The largest zone of quasi-periodic orbits spreads around the minimum of the effective potential $V$, where the guiding center approximation is valid \citep{1963RvGSP...1..283N}. Most studies of trapped particles have been focused on this region with relatively low energies. In Figure \ref{fig1a}, we note that for particles with initial conditions of $z=0$, $p_\rho=0$, and $\rho=1$, that is at the minimum of $V$, most orbits with $p_z>0$ are actually chaotic. Although these orbits have a very low energy, the guiding center approximation is invalid since these particles move initially along the magnetic field line at the equatorial plane \citep{doi:10.1063/1.859308}.

To further demonstrate this feature, in the \{$\rho$, $p_\rho$\} plane we mark the positions of ten orbits of particles with the same energy each time they cross the equatorial plane upward, the so-called Poincar\'{e} map. These particles are injected from the equatorial plane at $\rho=1$ with their initial pitch angle separated by 0.1 in their sine starting from 0. Figure \ref{fig12} shows these maps for the energy levels of 1/1000, 1/250, 1/100, and 1/50. These figures are similar to those in \citet{2016PhRvE..94d3203S}, where the energy levels are given in physical units instead of the dimensionless units here. Particles with larger pitch angles correspond to larger loops in the Poincar\'{e}. It is interesting to note that even in the case of the lowest energy, the Poincar\'{e} map deviates from the simple loop geometry for quasi-periodic orbits for the inner most particle injected along the magnetic field line (top-left panel). Its Poincar\'{e} map has a belt like structure at the beginning indicating a low value of the maximum Lyapunov exponent. It takes the reciprocal of the maximum Lyapunov exponent, the so-called Lyapunov time, for the orbit to become chaotic. The other orbits are obviously quasi-periodic at the lowest energy of 1/1000. 

{\bf However, the Poincar\'{e}e maps of these quasi-periodic orbits do not appear to be associated with a stable periodic orbit. The orbit in the equatorial plane is an apparent stable periodic orbit at the same energy. However, it does not have fixed points in the Poincar\'{e} map defined above. If it is plotted in the \{$\rho$, $p_\rho$\} plane, it encloses all points of the Poincar\'{e} map at the same energy. If one makes Poincar\'{e} maps in the plane of \{$z$, $p_z$\} with $\rho=1$ and $p_\rho>0$, orbits in the equatorial plane have a fixed point at the origin. The quasi-periodic orbits discussed above will have a Poincar\'{e} map surrounding this fixed point. One should note that due to the presence of unstable periodic orbits \citep{1990CeMDA..49..327J}, between contours of quasi-periodic orbits at a given energy, there are fixed points of unstable periodic orbits \citep{doi:10.1002/cpa.3160260204}. However, these unstable periodic orbits has zero measure in the phase space and may not be realized in nature.} 

In other high-energy levels, we see that orbits become chaotic even for particles with relatively large initial pitch angles, which agrees with Figure \ref{fig1a}. In the top-right panel of Figure \ref{fig12}, there appears to be crescent shape empty regions allowed by energy conservation even for chaotic orbits. {\bf The two lower panels suggest that these empty regions are likely associated with quasi-periodic orbits there. The finite volume of quasi-periodic orbits in the 4D phase space prevents chaotic trajectories from entering zones of quasi-periodic orbits. Therefore chaos does not imply a uniform distribution in the phase space allowed by energy conservation. Actually, one can see that chaotic orbits with relatively small pitch angles will not run into the regime of quasi-periodic orbits with large pitch angles. 
The lower panels also suggest a transition from quasi-periodic orbits to chaotic orbits when Poincar\'{e} maps of some orbits show crescent shape, reminiscence of bifurcation in dynamical systems.}

For particles that are static at the beginning with $p_z=p_\rho=0$, they will always be static if they are injected at the minimum of $V$ where $\dot{\phi}=0$. They correspond to static particles in the 3D real space. For $V>0$, these particles actually move in the toroidal direction $ \hat{\phi}$ at the beginning. Figure \ref{fig1b} shows that, near the polar regions, the maximum Lyapunov exponent of static particles can still be positive. This is because that in these regions, the initial points next to these static particles have chaotic orbits. Although the static motion itself is relatively simple, slight changes in the initial conditions will lead to chaos, giving rise to a positive Lyapunov exponent. The same is true for unstable periodic orbits \citep{devogelaere1950}.  The static particle at the saddle point of $z=0$ and $\rho=2$ corresponds to circular orbits in the equatorial plane with the centrifugal force balanced by the Lorentz force.

\begin{figure}[!htbp]
\includegraphics[width=0.9\columnwidth, angle=0.0]{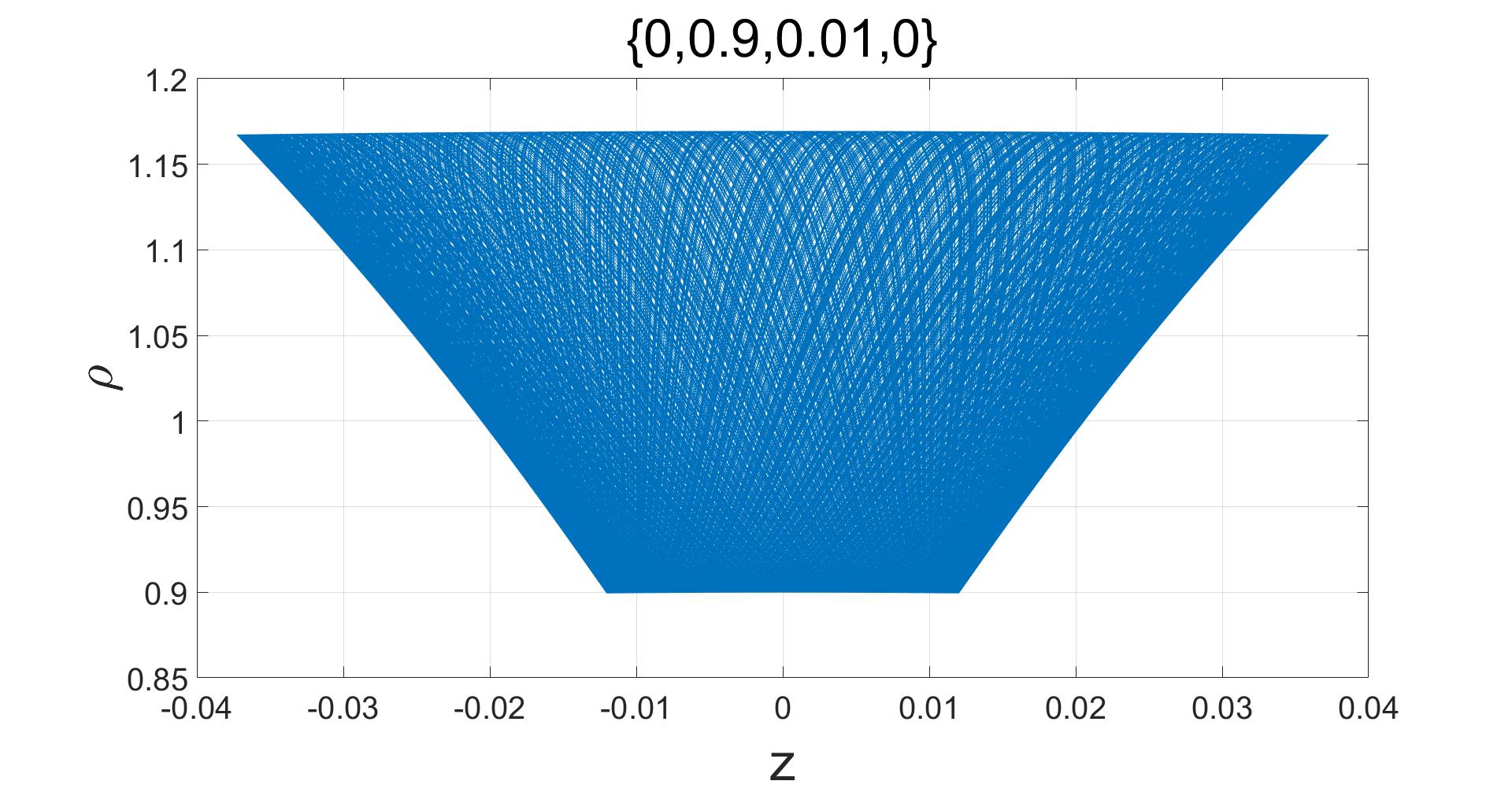}\\
\includegraphics[width=0.8\columnwidth, angle=0.0]{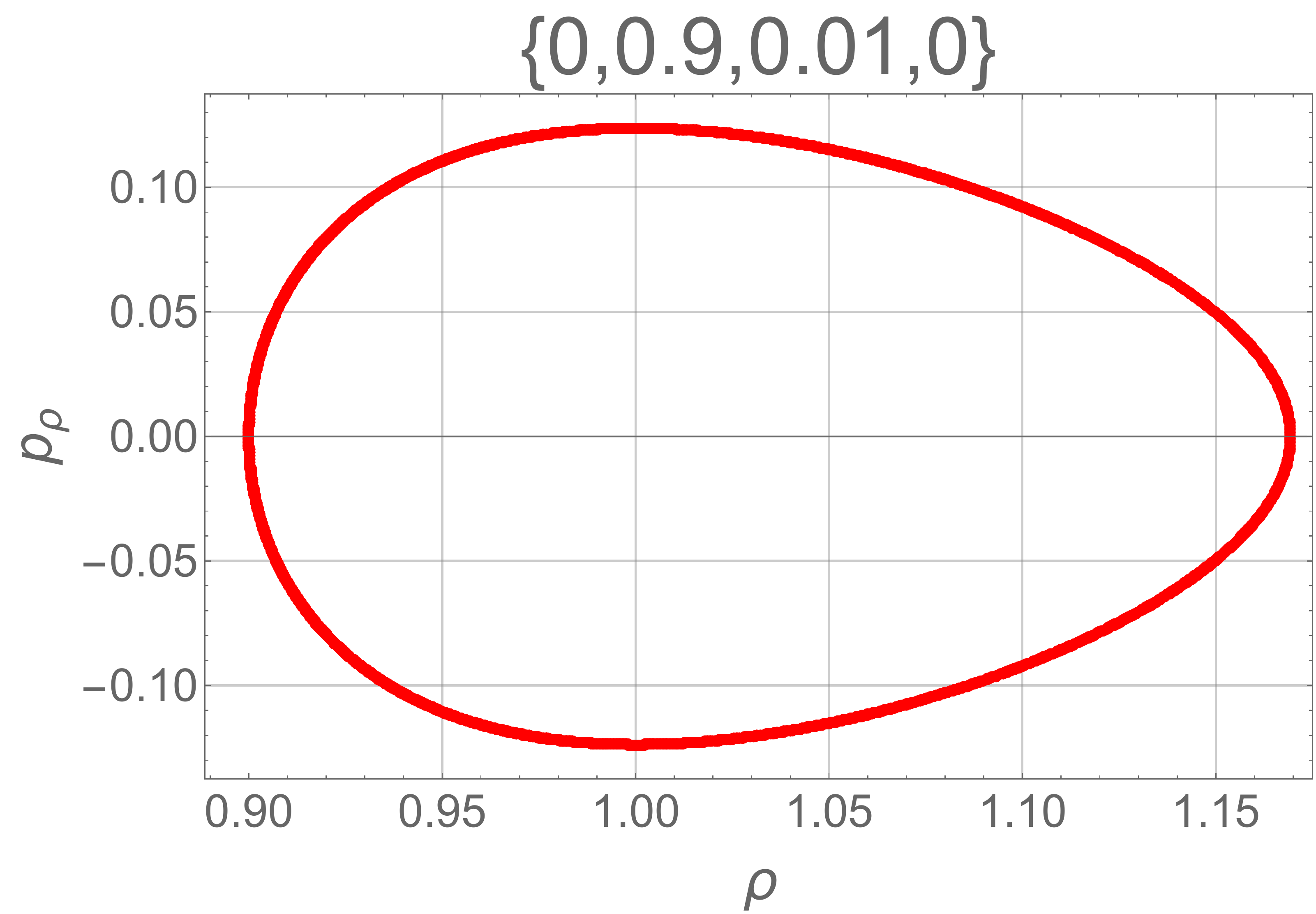}
\caption{\label{fig2} Typical quasi-periodic orbit (top) and Poincar\'{e} map (bottom) of a low energy (h $=0.0076708$) particle  with the initial conditions of \{0, 0.9, 0.01, 0\} indicated above the figure.}
\end{figure}

\begin{figure}[!htbp]
\centering
\includegraphics[width=0.9\columnwidth, angle=0.0]{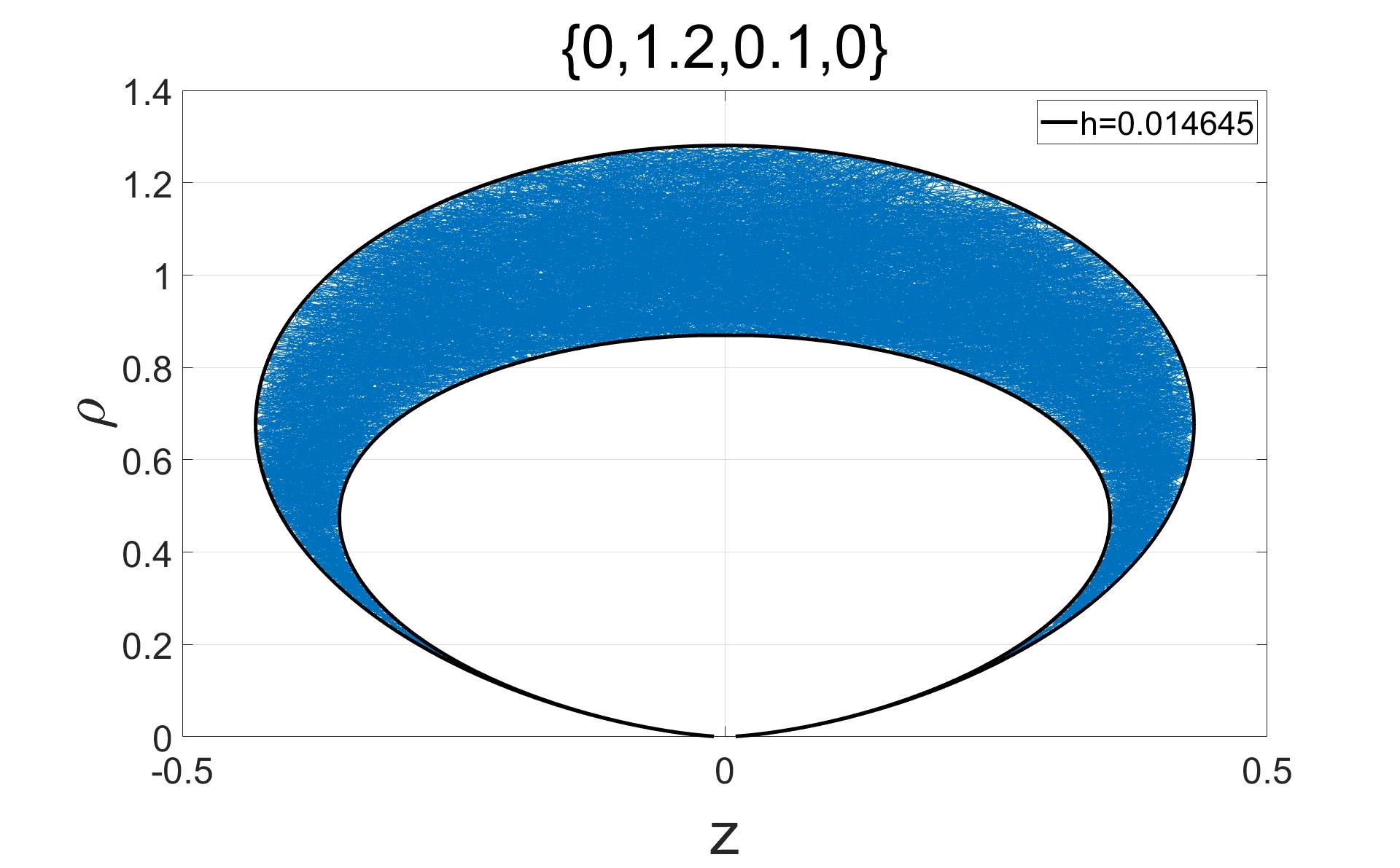}\\
\includegraphics[width=0.8\columnwidth, angle=0.0]{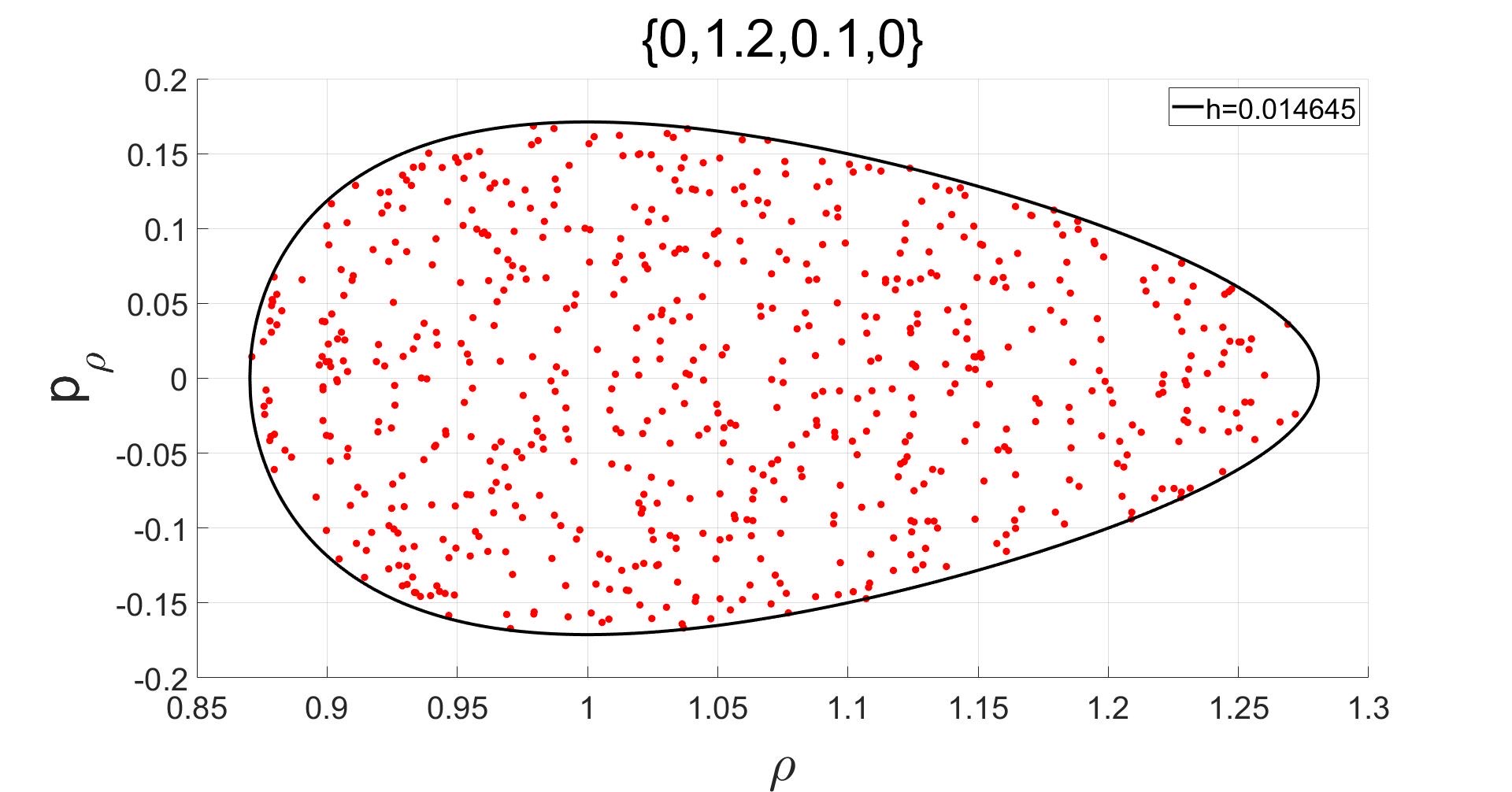}
\caption{\label{fig3} Typical chaotic orbit (top) and Poincar\'{e} map (bottom) of a low energy (h $=0.014645$) particle with the initial condition of \{0, 1.2, 0.1, 0\} indicated above the figure. The solid lines indicate the corresponding energy contours.}
\end{figure}

{\bf Both Figure \ref{fig01} and the right panel of Figure \ref{fig1c} show that the maximum Lyapunov exponent has a continuous distribution, implying gradual transition from quasi-periodic to chaotic motions. Simulations by \citet{2016PhRvE..94d3203S} and Figure \ref{fig12} also shows that the Poincar\'{e} maps can evolve continuously from points distributed along a ring to those in a belt and eventually spreading over the phase space constrained by the energy conservation. Nevertheless, sets of quasi-periodic orbits need to be open due to their vanishing Lyapunov exponents. Moreover, orbits with relatively low Lyapunov exponents can have properties similar to those of the quasi-period orbits \citep{2018PhRvL.121w5003H} within the Lyapunov time and may have significant implications on high energy electron observations in space \citep{2014Natur.515..531B}.}

To further test the robustness of the above results, we plot trajectories of a few orbits in the \{$\rho$, $z$\} plane and their Poincar\'{e} maps. Figure \ref{fig2} corresponds to a typical quasi-periodic orbit with a relatively low energy with $\{z, \rho, p_z, p_\rho\}=\{0, 0.9, 0.01, 0\}$ initially. The trajectory in the \{$\rho$, $z$\} plane follows a very regular pattern and the Poincar\'{e} map is a closed loop, both of which are defining characteristics of quasi-periodic orbits.
Figure \ref{fig3} corresponds to a chaotic orbit with $z=0$, $\rho= 1.2$, $p_z=0.1$, and $p_\rho = 0$ initially. In contrast to the orbit in Figure \ref{fig2}, the trajectory in the \{$\rho$, $z$\} and the Poincar\'{e} map cover almost the whole region allowed by the energy conservation.
 
\begin{figure}[!htbp]
\includegraphics[width=0.93\columnwidth, angle=0.0]{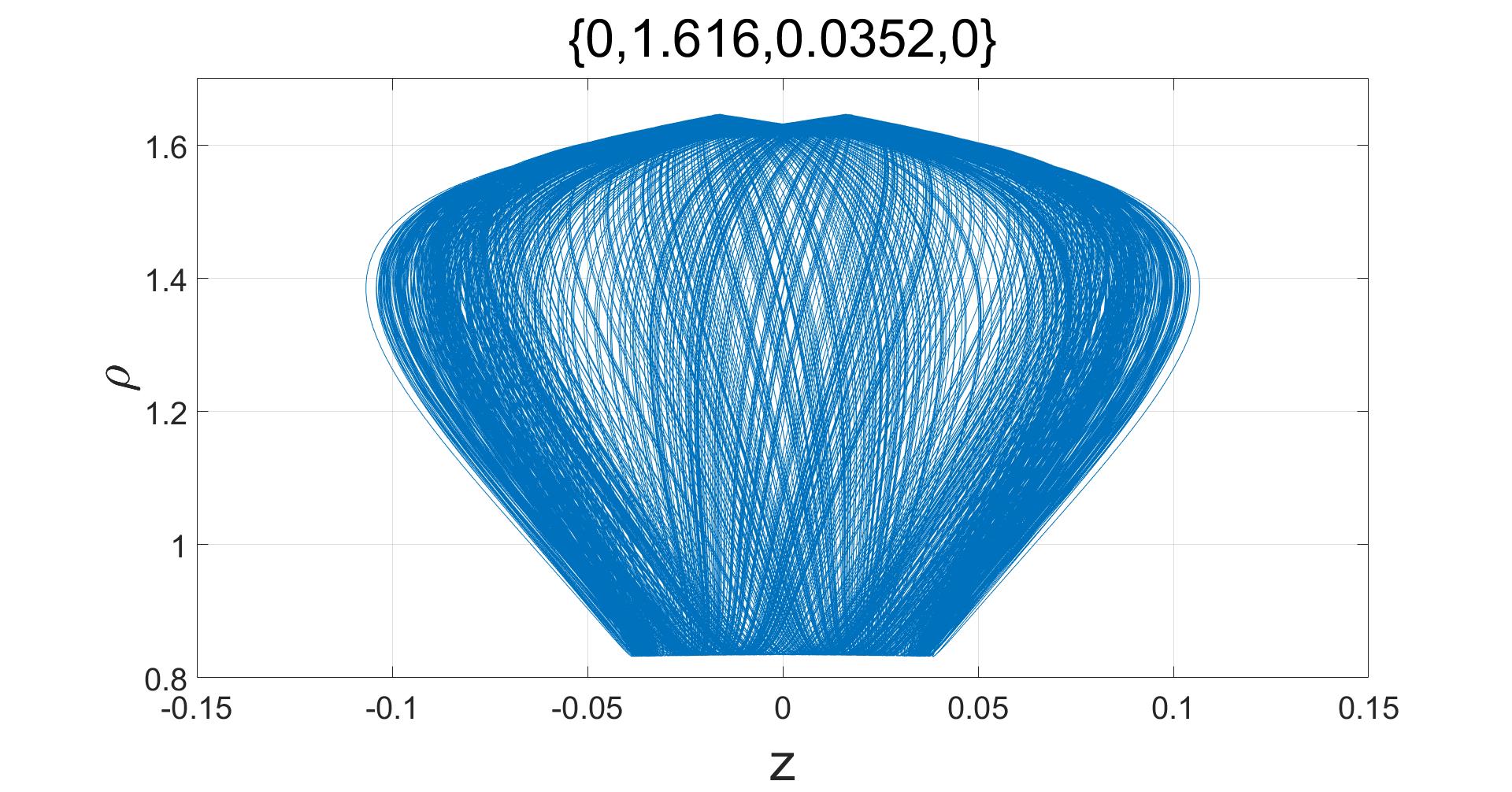}\\
\includegraphics[width=0.85\columnwidth, angle=0.0]{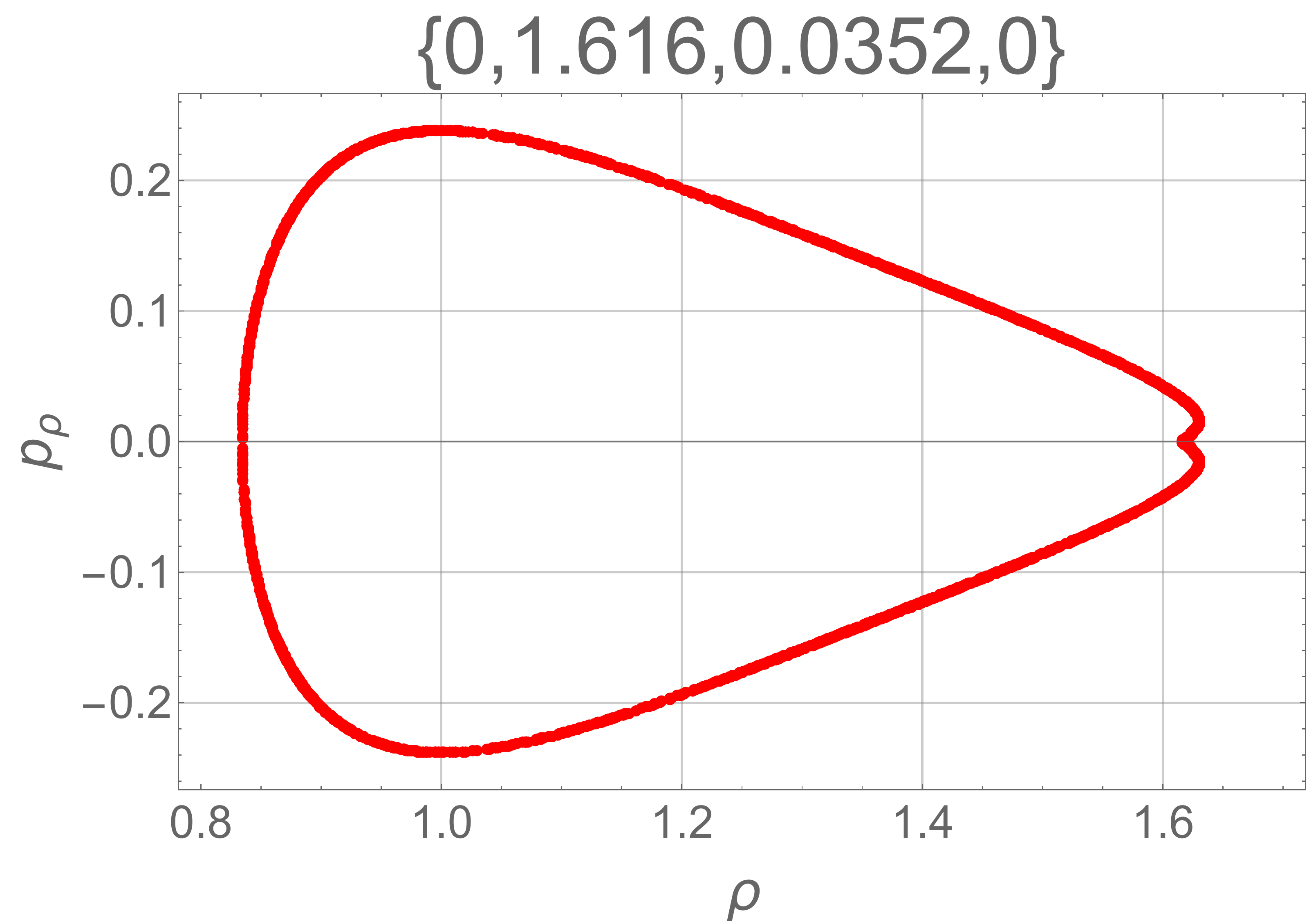}

\caption{\label{fig7} Typical quasi-periodic orbit (top) and Poincar\'{e} map (bottom) of a high energy (h $=0.02844$) particle with the initial condition of \{0, 1.616, 0.0352, 0\}  indicated above the figure.}
\end{figure}

\begin{figure}[!htbp]
\includegraphics[width=0.93\columnwidth, angle=0.0]{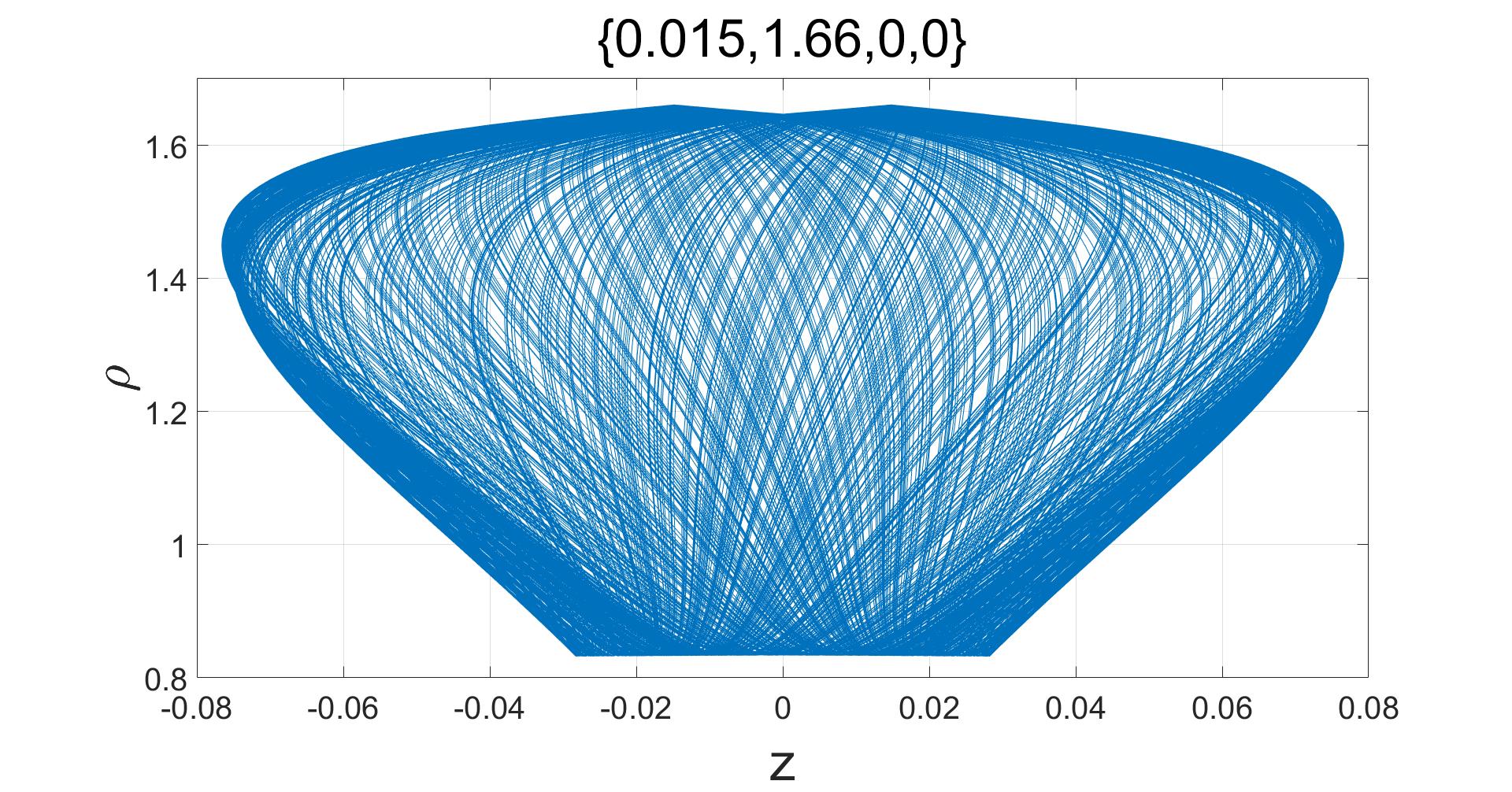}\\
\includegraphics[width=0.85\columnwidth, angle=0.0]{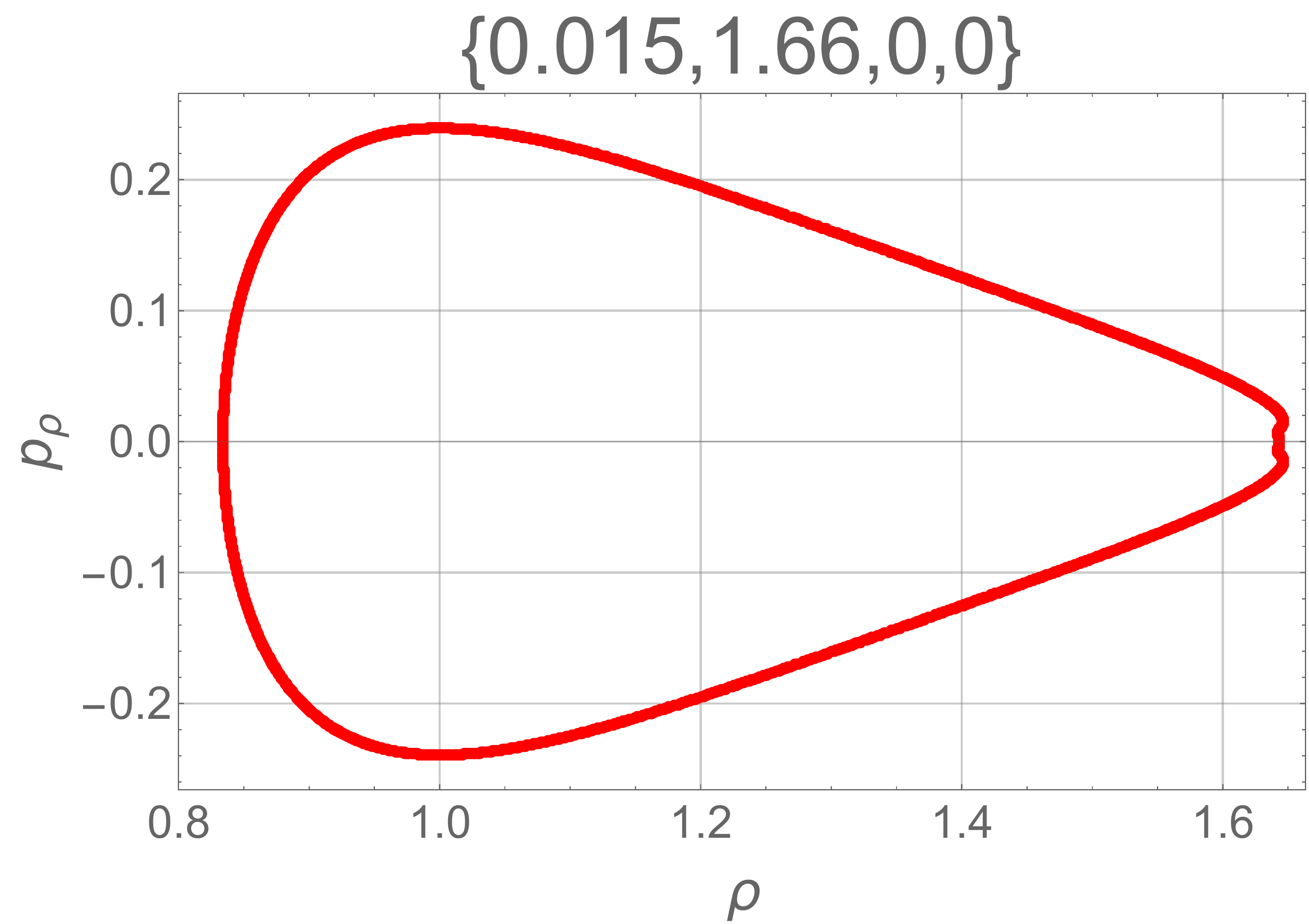}
\caption{\label{fig8} Typical quasi-periodic orbit (top) and Poincar\'{e} map (bottom) of a high energy (h $=0.028694$) particle with the initial condition of \{0.015, 1.66, 0, 0\}  indicated above the figure.}
\end{figure}

\begin{figure}[!htbp]
\includegraphics[width=0.93\columnwidth, angle=0.0]{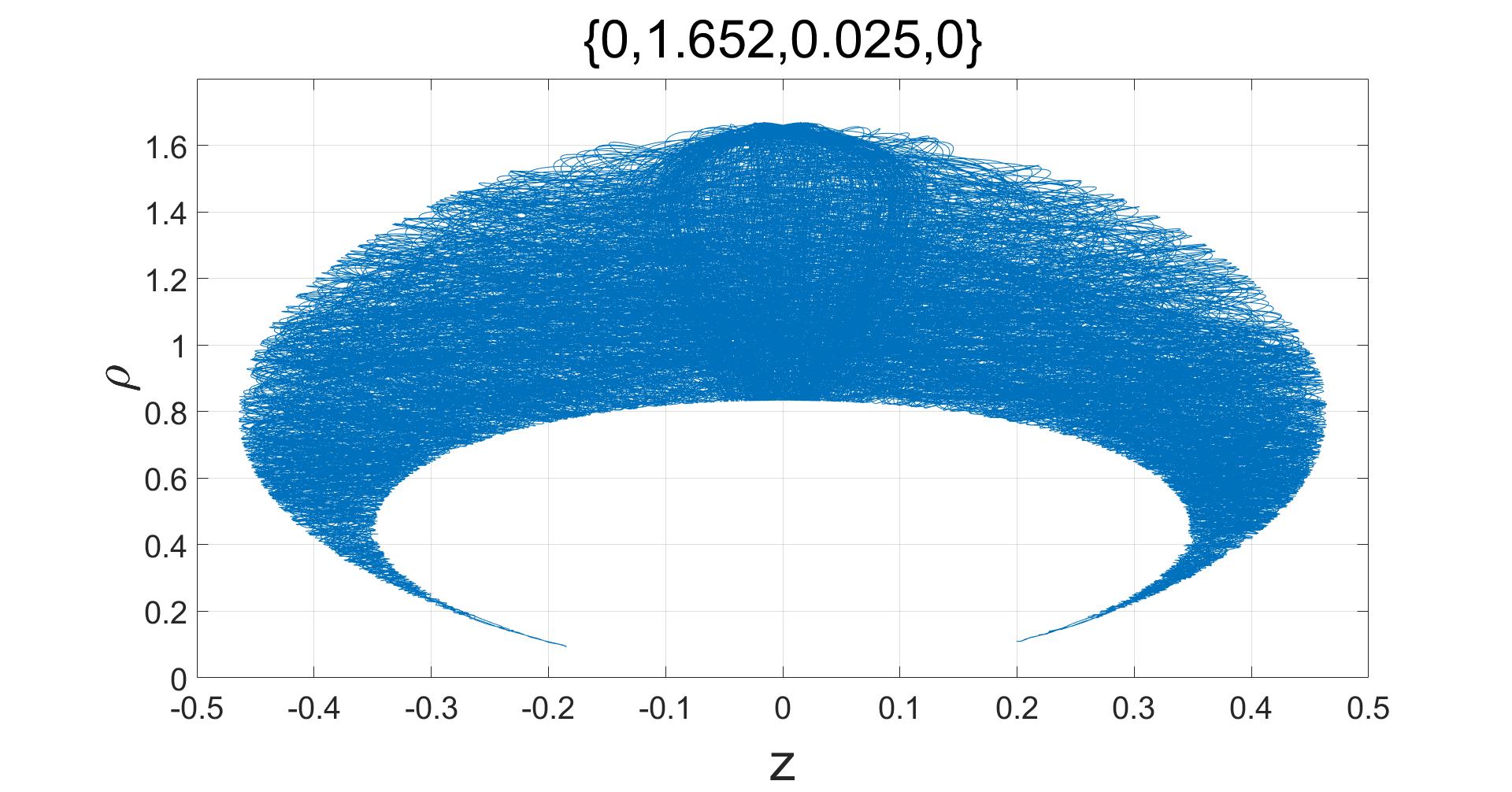}\\
\includegraphics[width=0.85\columnwidth, angle=0.0]{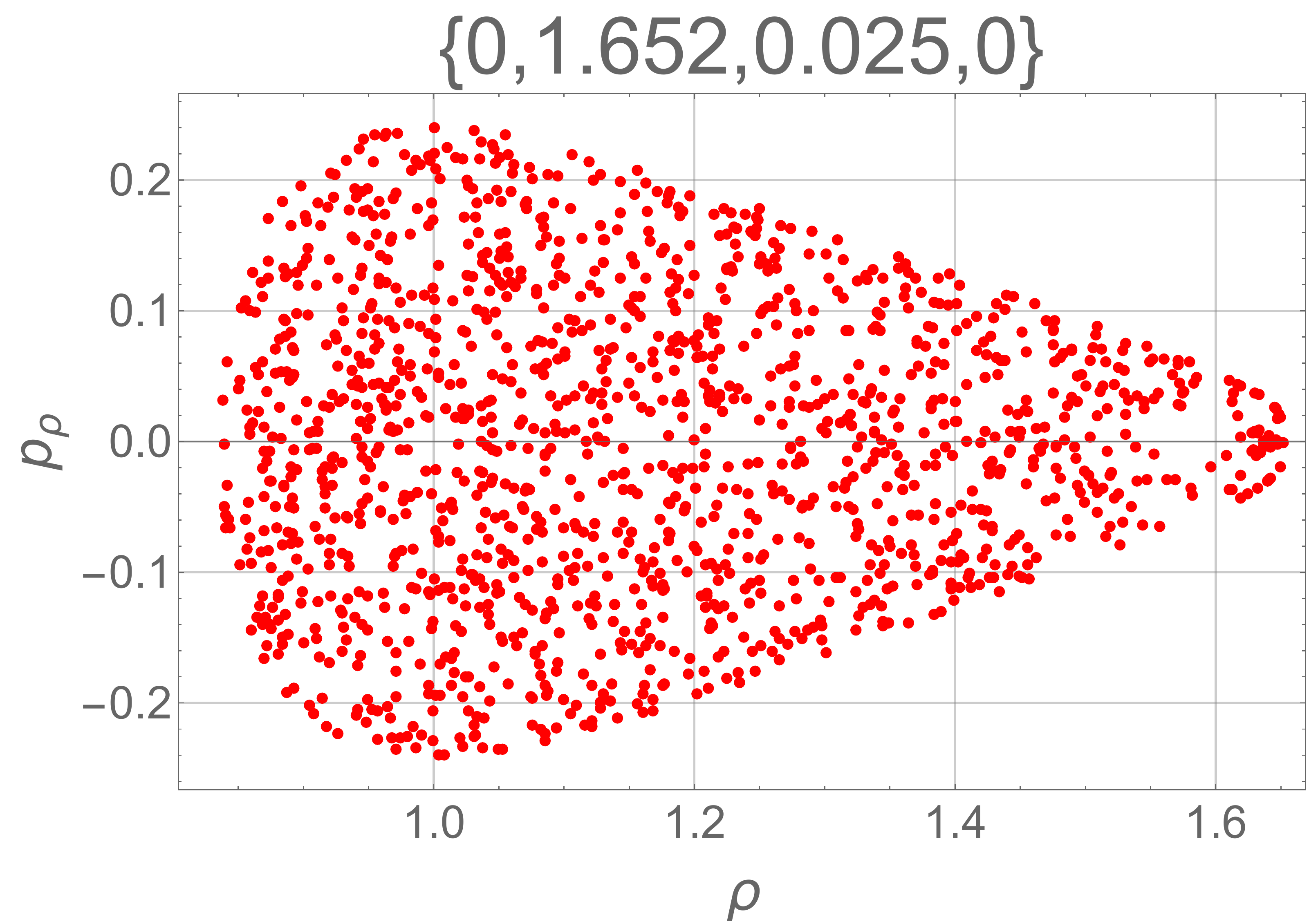}

\caption{\label{fig9} Typical chaotic orbit (top) and Poincar\'{e} map (bottom) of a high energy (h $=0.028851$) particle with the initial condition of \{0, 1.652, 0.025, 0\}  indicated above the figure.}
\end{figure}

Figure \ref{fig7} represents a typical quasi-periodic orbit in the first set with $z=0$, $\rho= 1.616$, $p_z=0.0352$, and $p_\rho = 0$ initially. Figure \ref{fig8} is similar with a different set of initial conditions: $z=0.015$, $\rho= 1.66$, $p_z=0.0$, and $p_\rho = 0$. The Poincar\'{e} maps are expected to be a close loop for long enough period of calculations. The cave-in of these loops at the high value end of $\rho$ appears to be associated with the presence of the first set of quasi-periodic orbits there. Figure \ref{fig9} shows a typical chaotic orbit and its Poincar\'{e} map of a particle with $z=0$, $\rho= 1.652$, $p_z=0.025$, and $p_\rho = 0$ initially, which is surrounded by the first set of quasi-periodic orbits.

\begin{figure}[!htbp]
\includegraphics[width=0.936\columnwidth, angle=0.0]{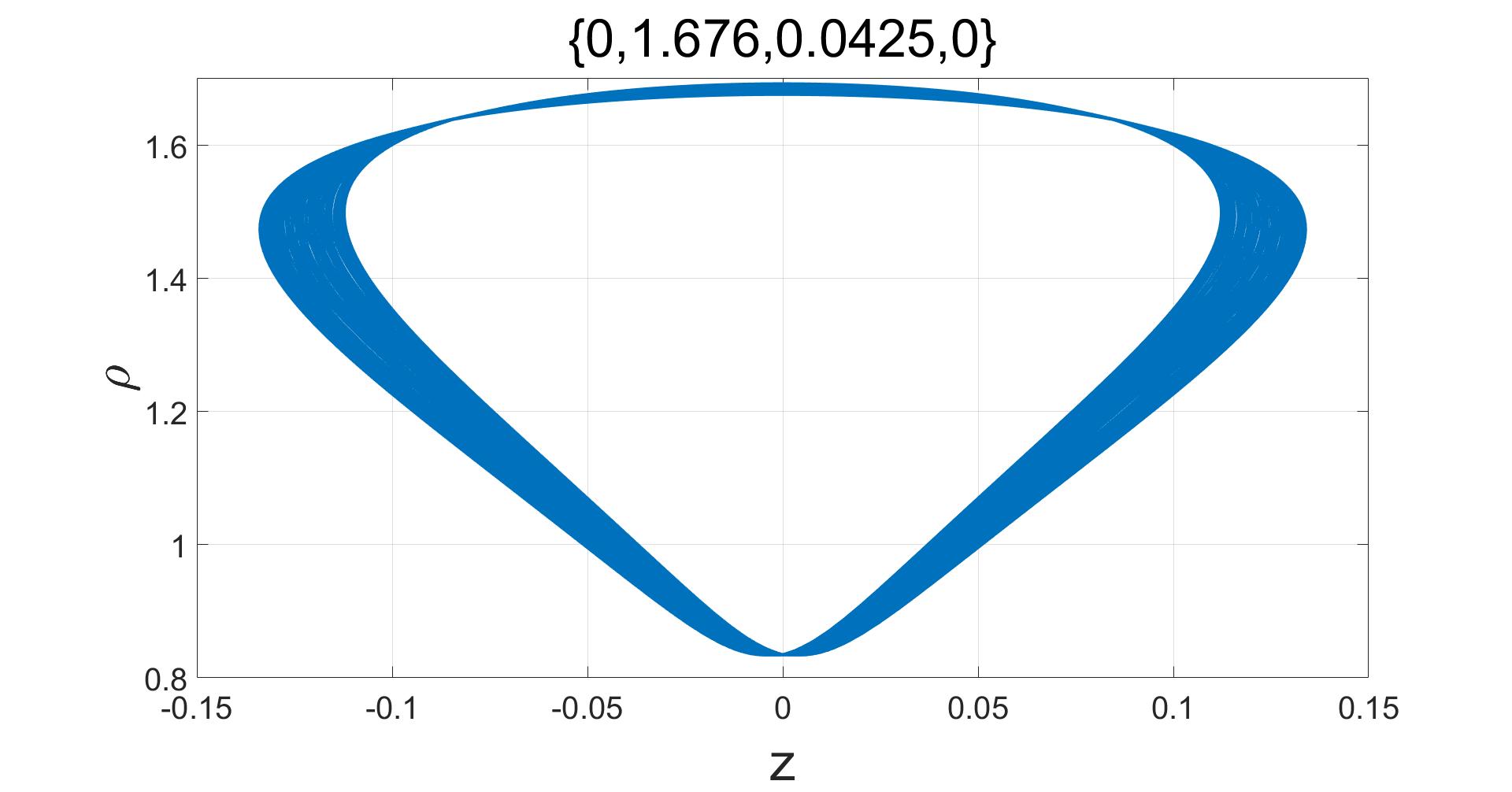}\\
\includegraphics[width=0.8\columnwidth, angle=0.0]{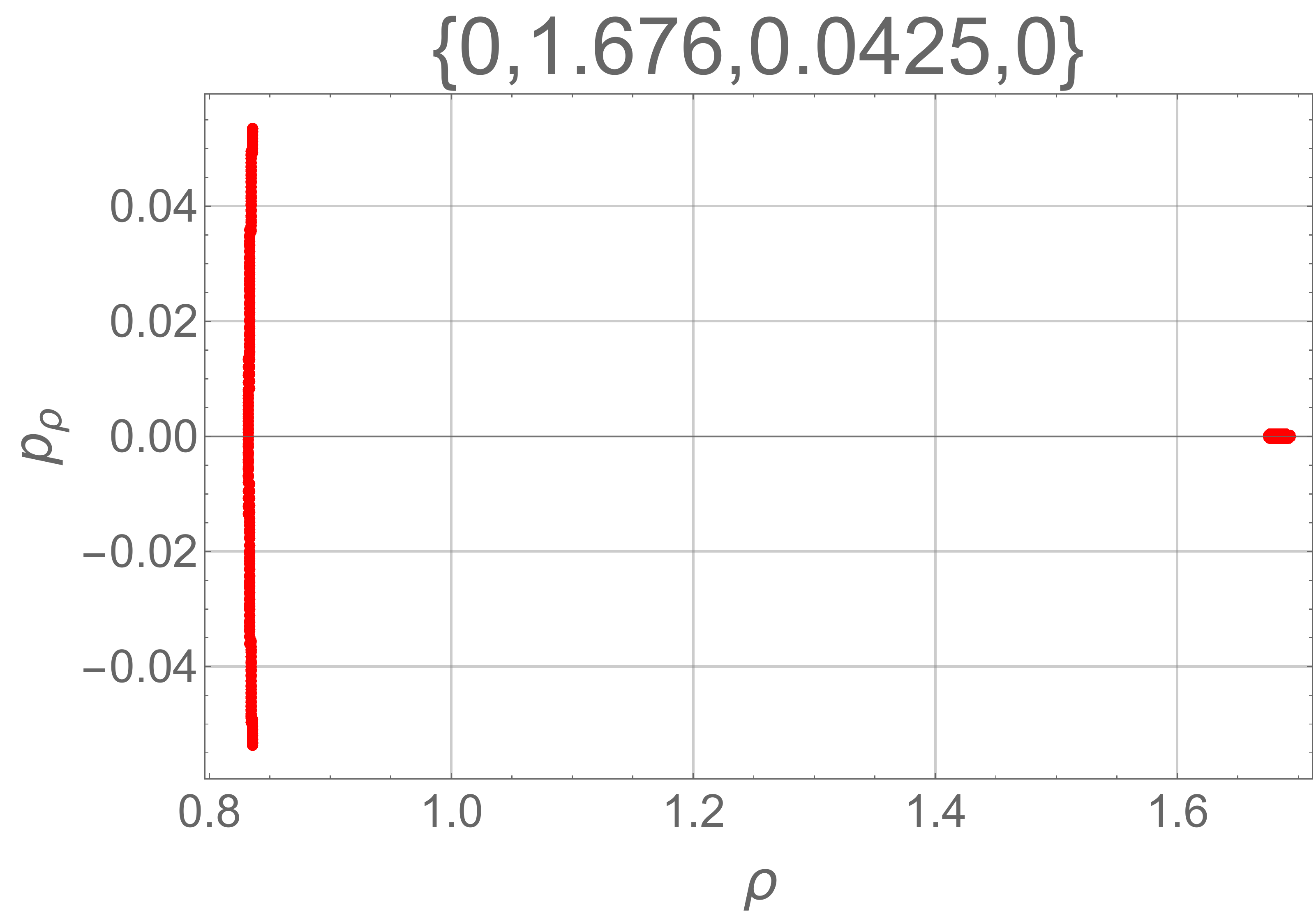}
\caption{\label{fig4} Typical quasi-periodic orbit (top) and Poincar\'{e} map (bottom) of a high energy (h $=0.029861$) particle with the initial condition of \{0, 1.676, 0.0425, 0\} indicated above the figure.}
\end{figure}

\begin{figure}[!htbp]
\includegraphics[width=0.936\columnwidth, angle=0.0]{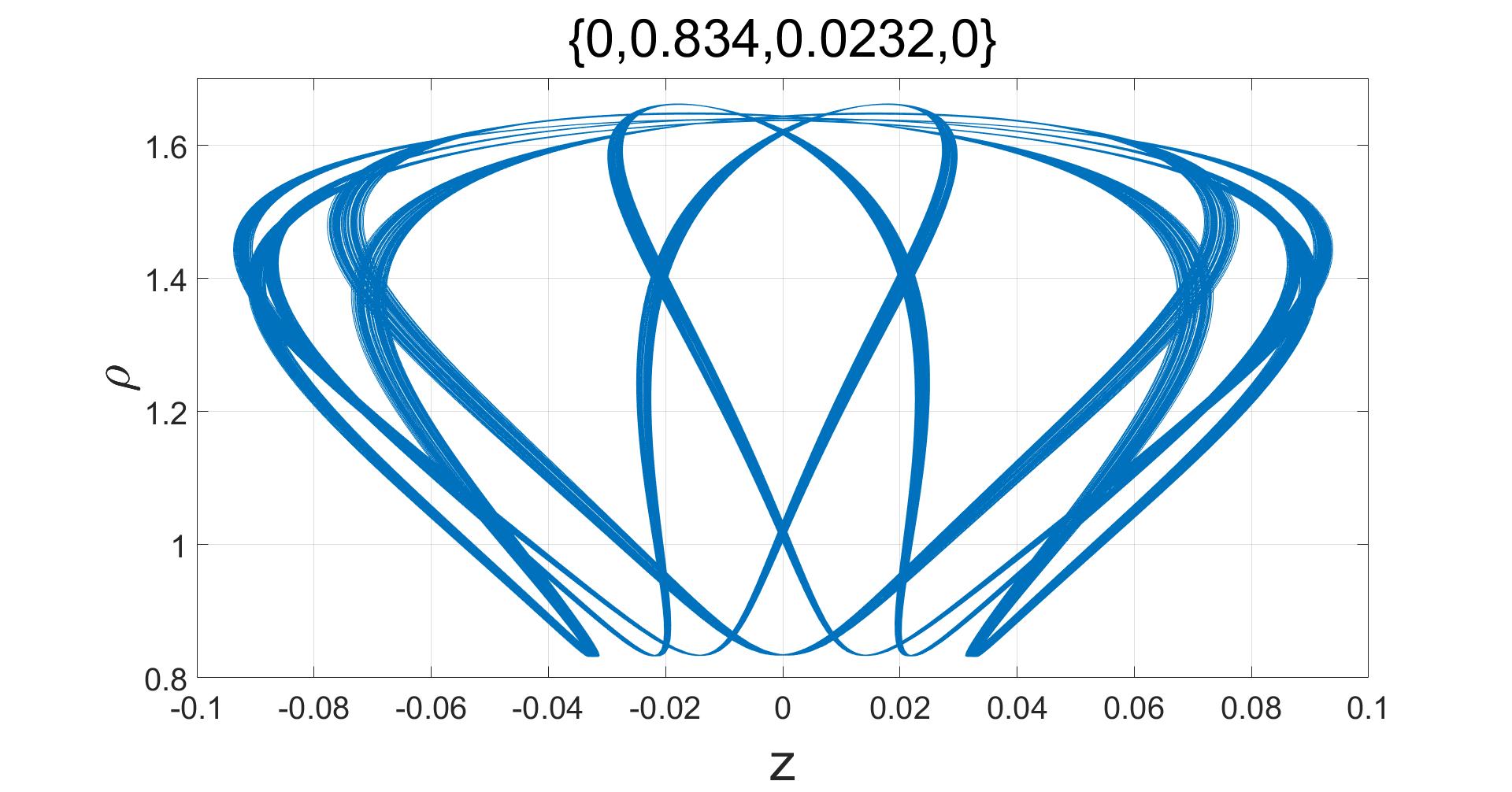}\\ 
\includegraphics[width=0.83\columnwidth, angle=0.0]{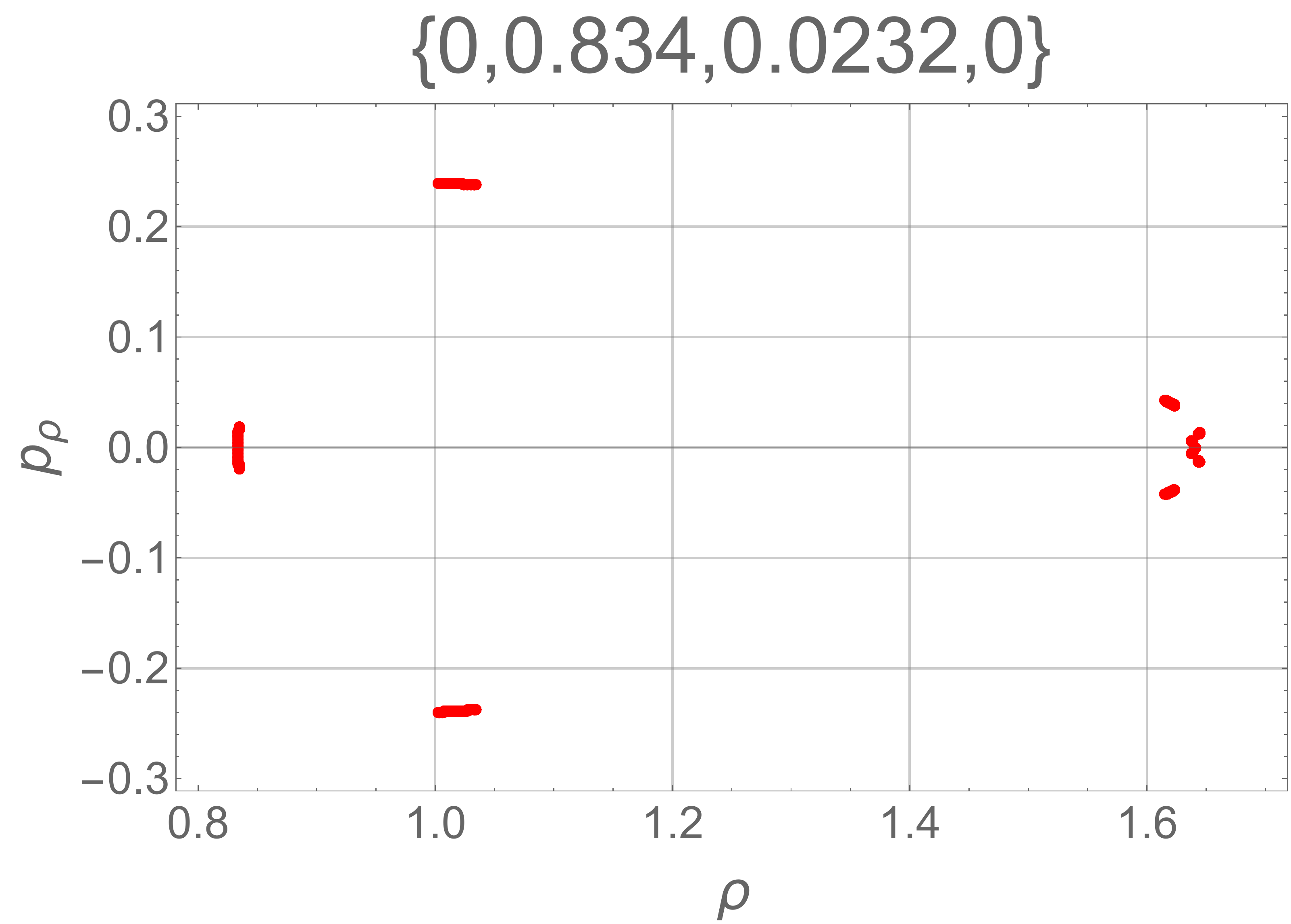}
\caption{\label{fig5} Typical quasi-periodic orbit (top) and Poincar\'{e} map (bottom) of a high energy (h $=0.028748$) particle with the initial condition of \{0, 0.834, 0.0232, 0\} indicated above the figure.}
\end{figure}

\begin{figure}[!htbp]
\includegraphics[width=0.93\columnwidth, angle=0.0]{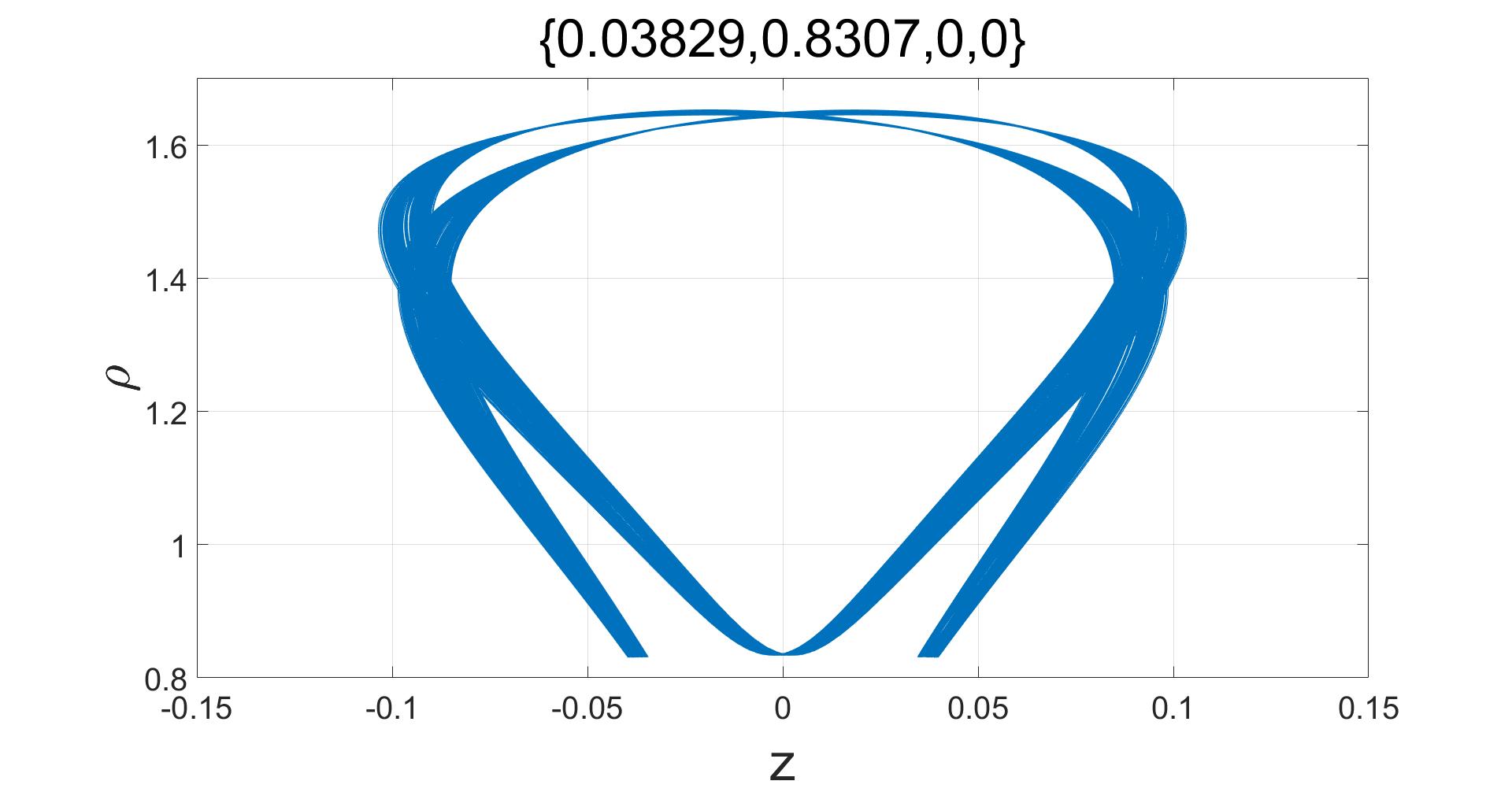}\\
\includegraphics[width=0.85\columnwidth, angle=0.0]{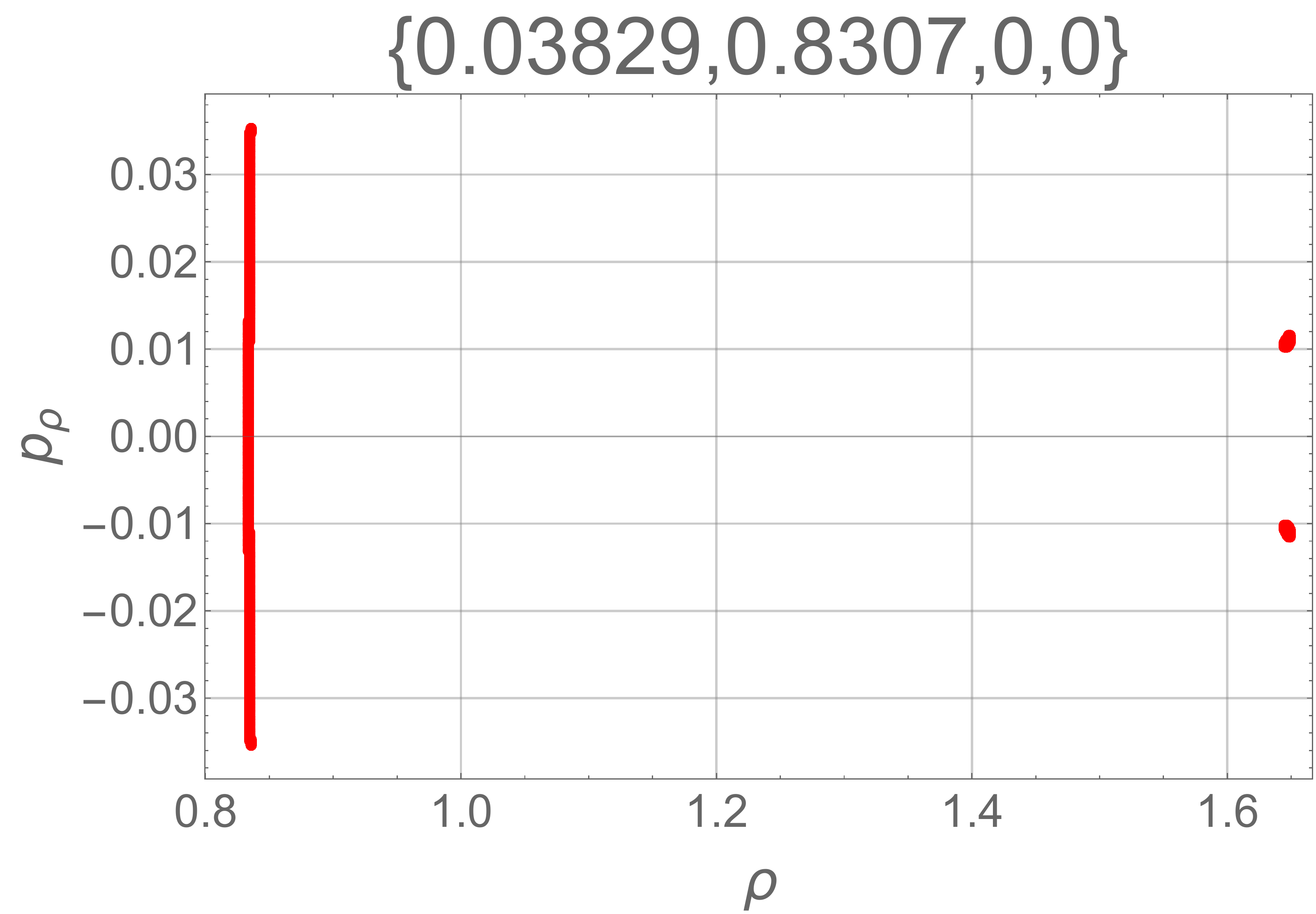}
\caption{\label{fig6} Typical quasi-periodic orbit (top) and Poincar\'{e} map (bottom) of a high energy (h $=0.028976$) particle with the initial condition of \{0.03829, 0.8307, 0, 0\} indicated above the figure.}
\end{figure}

Figure \ref{fig4} corresponds to another quasi-periodic orbit in the first set with a relatively high energy with $z=0$, $\rho= 1.676$, $p_z=0.0425$, and $p_\rho = 0$ initially. The Poincar\'{e} map appears to consist of two line segments, which is unusual. According to \citet{1990ComPh...4..549S}, quasi-periodic orbits are confined on a torus in the 4D phase space, and their Poincar\'{e} maps should have infinite points along closed loops  while those of periodic orbits only have a few isolated points.
Similarly, Figures \ref{fig5} and \ref{fig6} represent respectively quasi-periodic orbits in the first set with $z=0$, $\rho= 0.834$, $p_z=0.0232$, $p_\rho = 0$, and $z=0.03829$, $\rho= 0.8307$, $p_z=0$, $p_\rho = 0$  initially. These Poincar\'{e} maps have more segments. {\bf As shown in the appendix, these segments actually are closed loops. These quasi-periodic orbits appear to be associated with stable periodic orbits in the Meridian plane that have more than one fixed points. Our numerical calculations therefore can also be used to search for stable periodic orbits in the phase space.}

\begin{figure}[!htbp]
\includegraphics[width=0.93\columnwidth, angle=0.0]{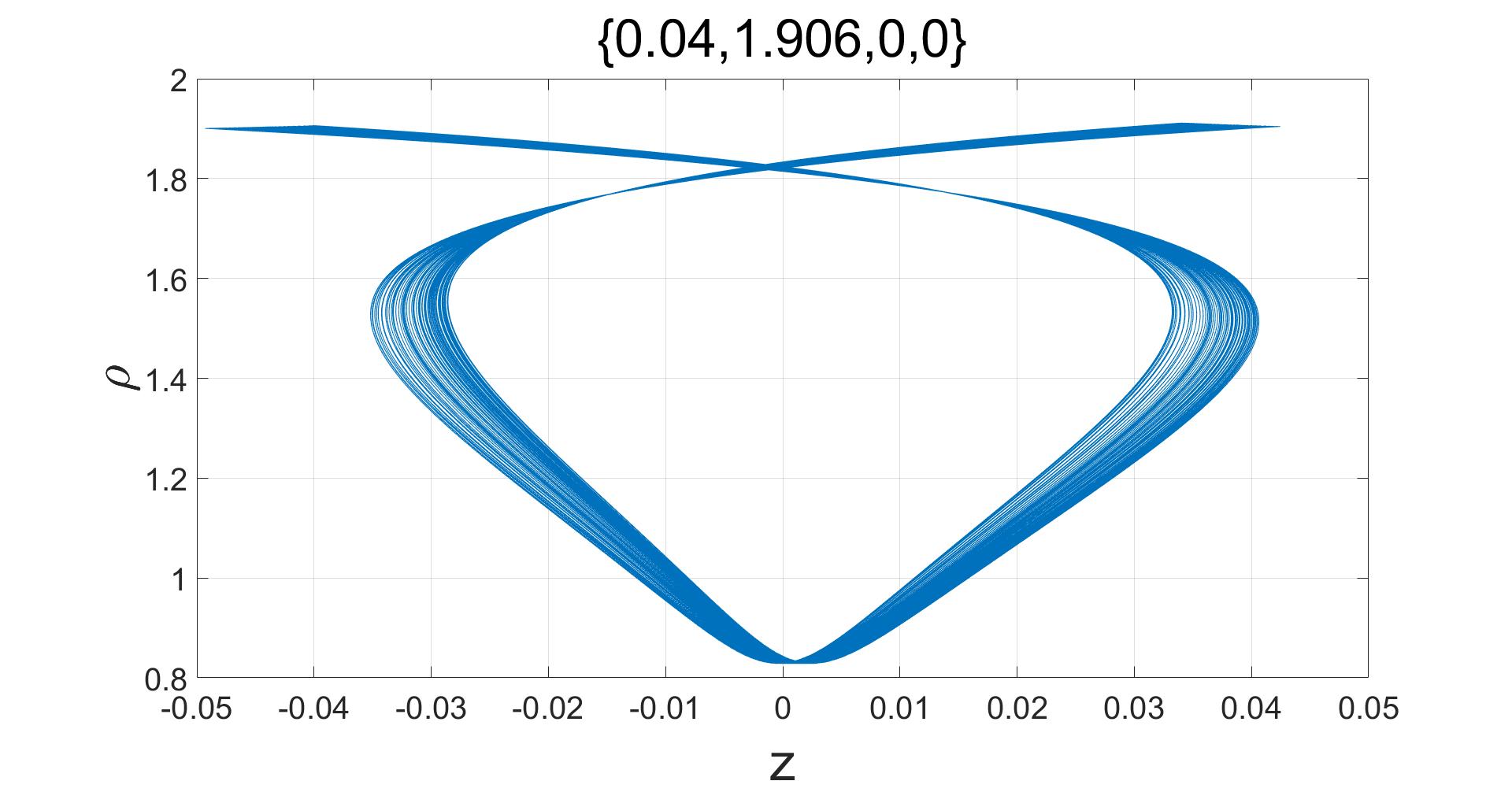}\\

\includegraphics[width=0.85\columnwidth, angle=0.0]{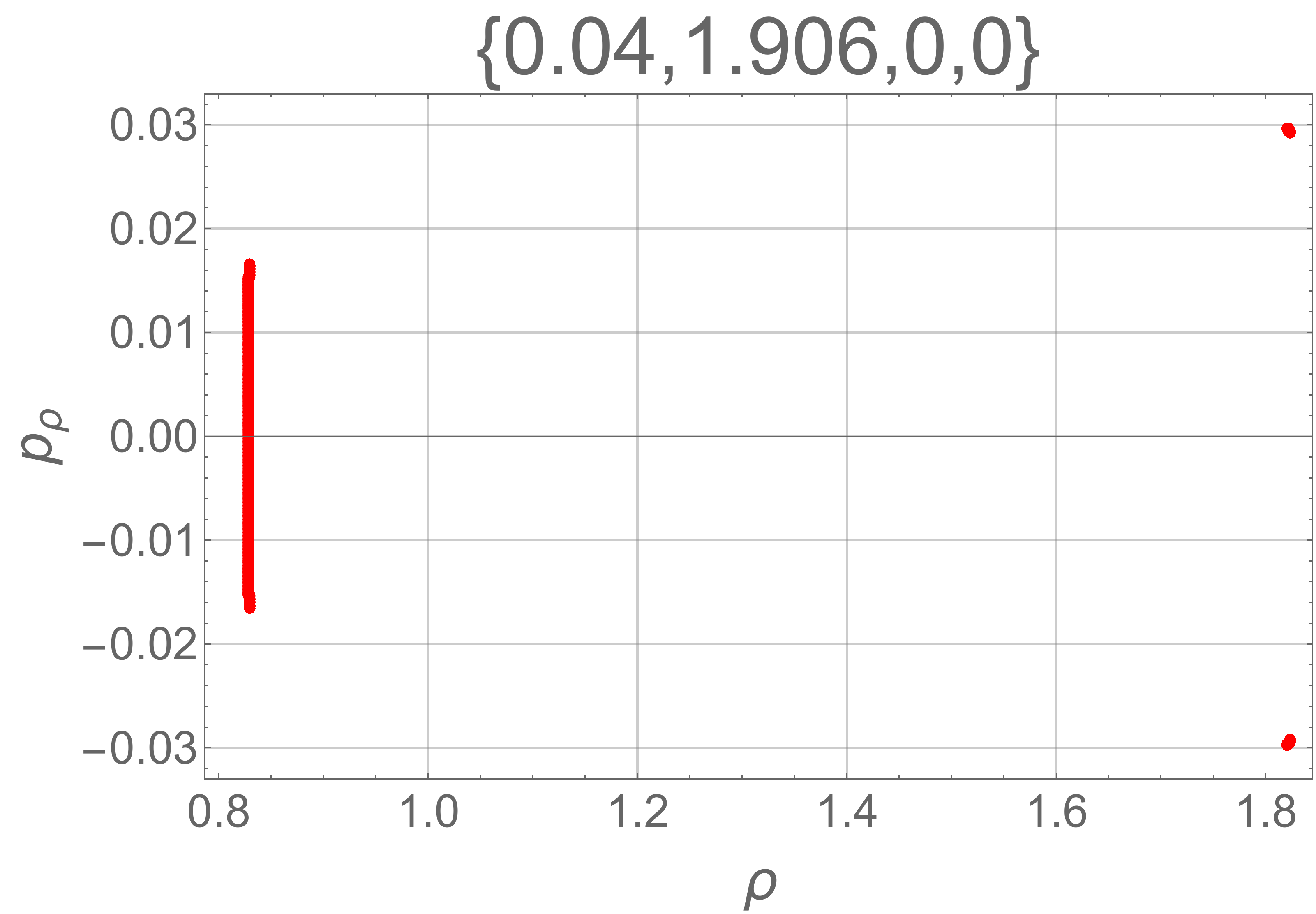}

\caption{\label{fig10} Typical quasi-periodic orbit (top) and Poincar\'{e} map (bottom) of a very high energy (h $=0.031144$) particle with the initial condition of \{0.04, 1.906, 0, 0\} indicated above the figure.}
\end{figure}

\begin{figure}[!htbp]
\includegraphics[width=0.93\columnwidth, angle=0.0]{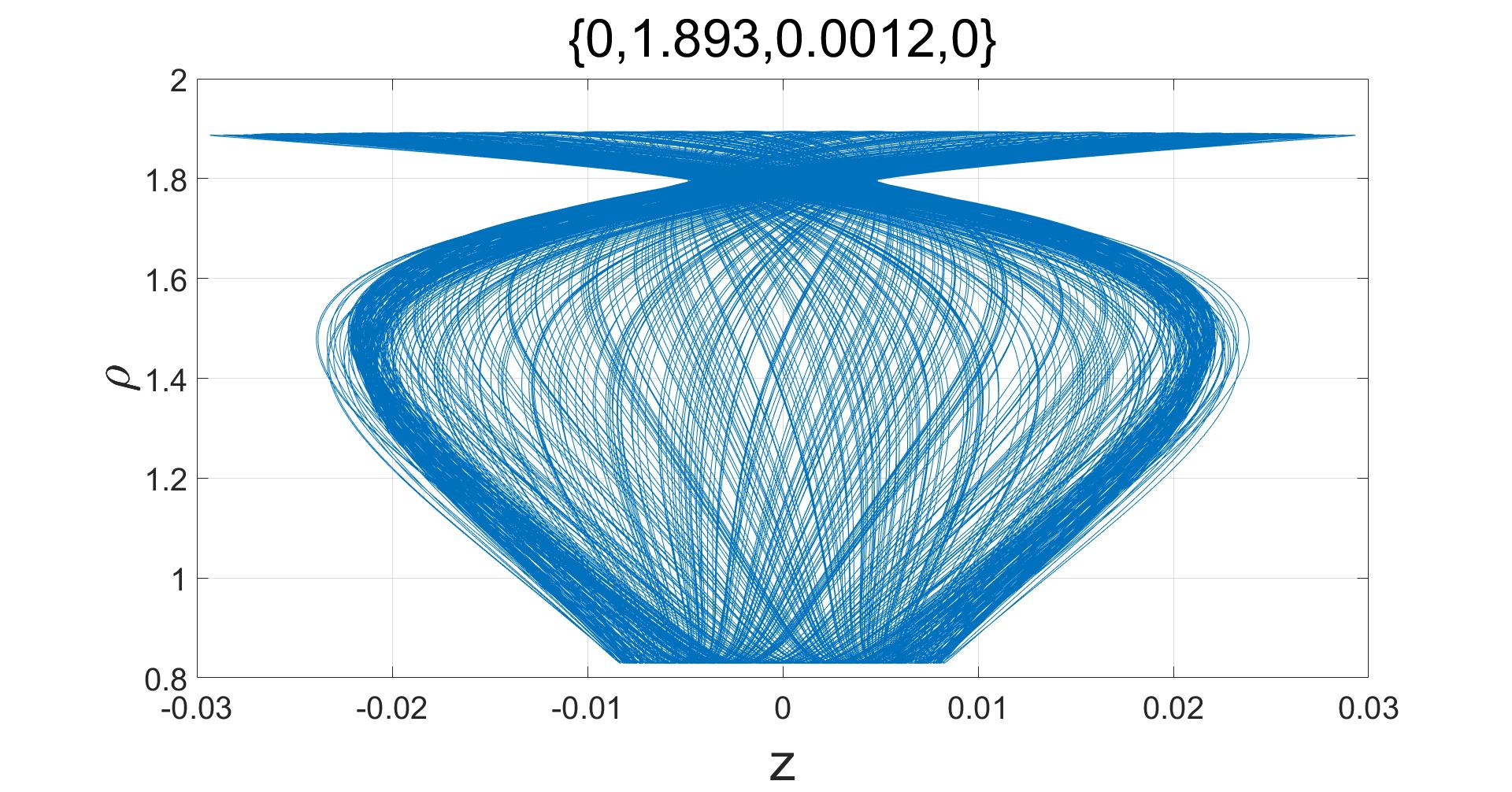}\\
\includegraphics[width=0.85\columnwidth, angle=0.0]{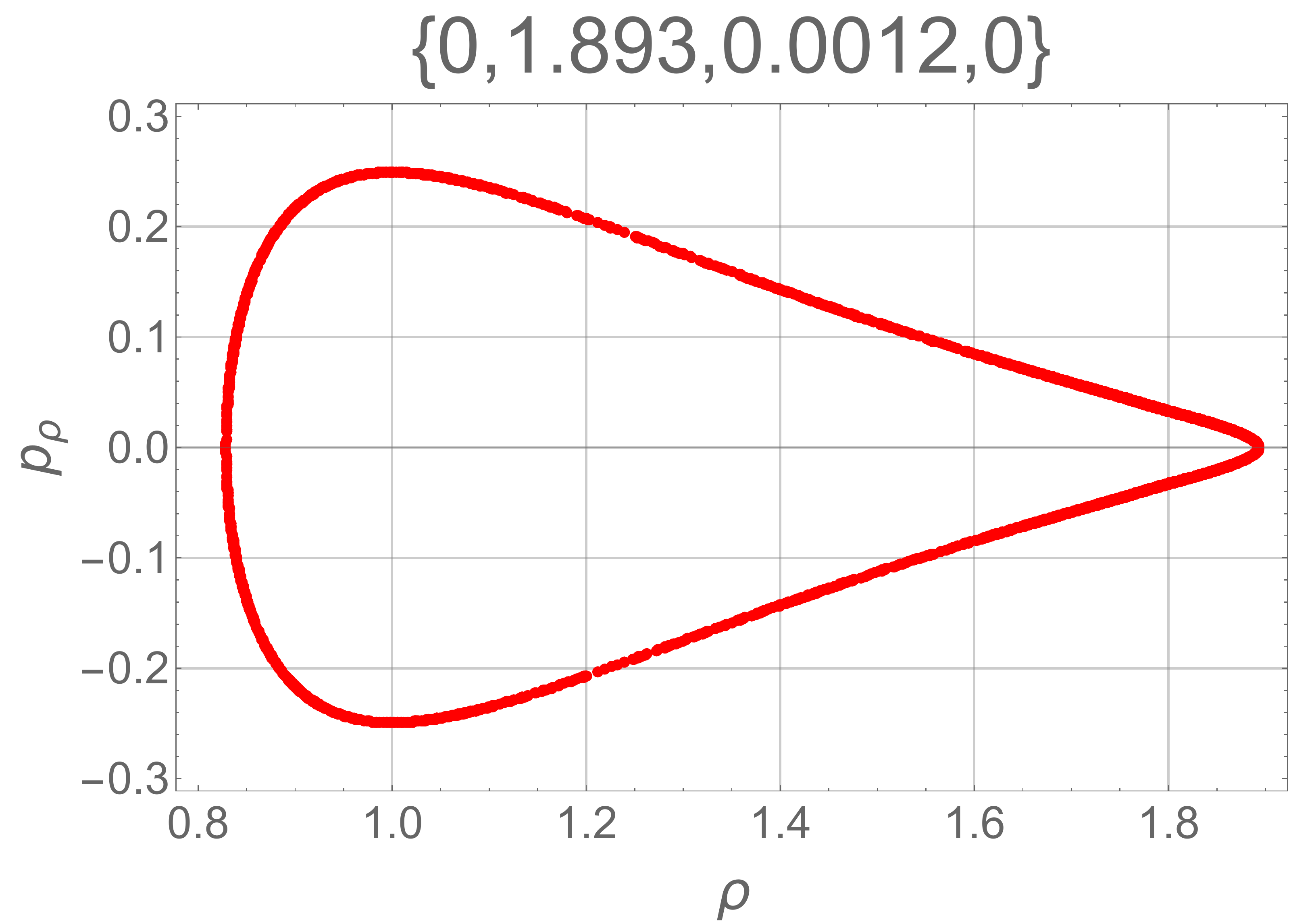}
\caption{\label{fig11} Typical quasi-periodic orbit (top) and Poincar\'{e} map (bottom) of a very high energy (h $=0.031051$) particle with the initial condition of \{0, 1.893, 0.0012, 0\} indicated above the figure.}
\end{figure}

Figures \ref{fig10} and \ref{fig11} show typical quasi-periodic orbits in the second set with an energy close to 1/32.2. Similar to Figure \ref{fig4}, the orbit in Figure \ref{fig10} is associated with stable periodic orbits in the Meridian plane.

\citet{devogelaere1950} first studied the stability of periodic orbits in the equatorial plane and found that there are infinite segments of stable orbits. The three sets of quasi-periodic orbits discovered here correspond to the three segments of stable orbits with the lowest energies. It can be anticipated that there are also sets of quasi-periodic orbits surrounding other segments of stable periodic orbits in the equatorial plane (See Figure \ref{fig1b}). However the volume of these sets are much smaller than the three sets discussed above so that most of them are not uncovered with our low resolution scans of the phase space.

\section{Conclusions}

The Lyapunov exponents measure the dependence of trajectory of a dynamical system on its initial conditions. Via evaluation of the Lyapunov exponents of charged particles trapped in a dipole magnetic field, we found prominent sets of quasi-periodic orbits with vanishing Lyapunov exponent around stable periodic orbits, in particular those in the equatorial plane \citep{doi:10.1142/S0218127400000177}. The volume of these sets of quasi-periodic orbits in the phase space of initial conditions appear to be significant. Particles in the low energy set have been extensively studied with the guiding center approximation \citep{1963RvGSP...1..283N}. Particles in the high energy sets should also be detectable in physical systems \citep{2014Natur.515..531B, NaokiKENMOCHI2019}. 

{\bf With the increase of energy, the particle orbits are more likely chaotic \citep{doi:10.1142/S0218127400000177}. However, the correlation between the maximum Lyapunov exponent and the energy are complicated. Our results also suggest that quasi-periodic orbits are always associated with stable periodic orbits. Both of them have vanishing Lyapunov exponents. Therefore one may use Poincar\'{e} maps of quasi-periodic orbits to search for stable periodic orbits in the phase space systematically. The maximum Lyapunov exponent of unstable periodic orbits, on the other hand, can be positive. Such unstable periodic orbits are usually difficult to identify \citep{1990ComPh...4..549S}. However, our results suggest that the maximum Lyapunov evolves continuously along segments of unstable periodic orbits, which may be used to identify branches of periodic orbits. 

Knowing that particles are making circular motion at the saddle point of the dimensionless potential, one can readily obtain the energy and location of particles in the quasi-periodic orbits discovered in this paper. In the case of the earth's magnetosphere, assuming an equatorial magnetic field of 0.25 G at the surface, the Lorentz factor of electrons making circular motion at 6 times the earth radius is $\gamma_{\rm max}\simeq 2.5\times 10^3$, which is much higher than those detected in the magnetosphere \citep{2014Natur.515..531B}. The corresponding electrons in the high-energy sets have an energy of about 1 GeV and should be oscillating around the equatorial plane in the radial range of 2.5 to 6 times the earth radius. Electrons in the low-energy set can have an energy up to 300 MeV and oscillate between 2.7 to 3.6 times the earth radius. The earth rotation and the misalignment of the earth magnetic field with respect to the rotation axis is expected to squeeze the volume of the quasi-periodic orbits. The KAM theorem however indicates that some quasi-periodic orbits should survive \citep{1963RuMaS..18....9A}. It is also possible that these perturbations may reduce the characteristic energy and make quasi-periodic orbits reach a much higher latitude, providing a possible explanation to the sharp cutoff of high energy electron distribution toward low altitude discovered by the Van Allen probes \citep{2014Natur.515..531B}. More realistic modelling of the earth's magnetic field is needed to clarify this issue.
}

\begin{acknowledgments}
We appreciate helpful discussions with Ms. Runna Wang and Mr. Anda Xiong. This work is partially supported by the National Key R\&D Program of China grant No. 2018YFA0404203, NSFC grants U1738122, U1931204, and 11761131007,  and by the International Partnership Program of the Chinese Academy of Sciences, grant No. 114332KYSB20170008.
\end{acknowledgments}

\section*{Data availability}

The data that support the findings of this study are available from the corresponding author upon reasonable request.\\[2ex]

\section*{Appendix}

\begin{figure}[!htbp]
\includegraphics[width=0.85\columnwidth, angle=0.0]{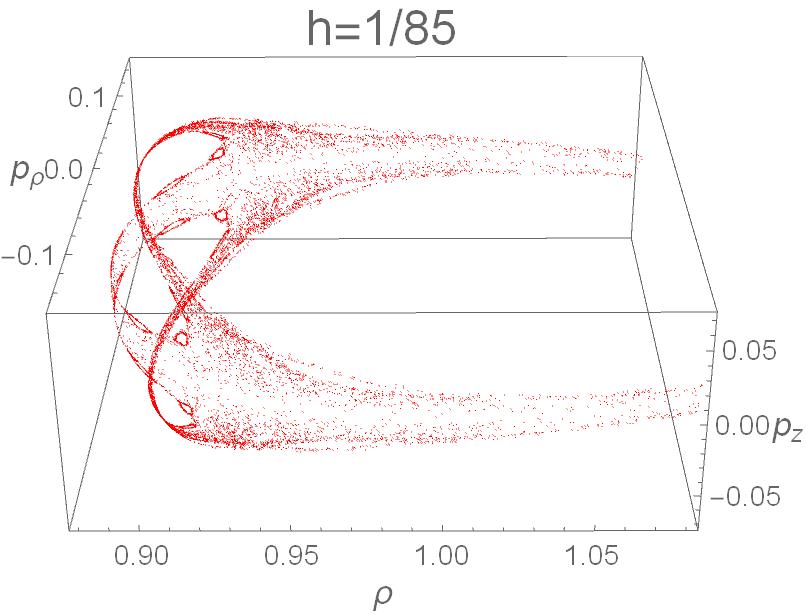}\\
\caption{\label{figa1} 3D Poincar\'{e} map of the outer particle shown in the bottom right panel of Figure \ref{fig12}.}
\end{figure}

{\bf Figure \ref{figa1} shows the 3D Poincar\'{e} map, i.e., the 3D distribution of points when the trajectory cross the equatorial plane, of a transition orbit between quasi-periodic and chaotic. Its 2D Poincar\'{e} map has a crescent shape as shown by the outer particle in the bottom right panel of Figure \ref{fig12}.}

\begin{figure*}[htbp]
\includegraphics[width=0.43\linewidth, angle=0.0]{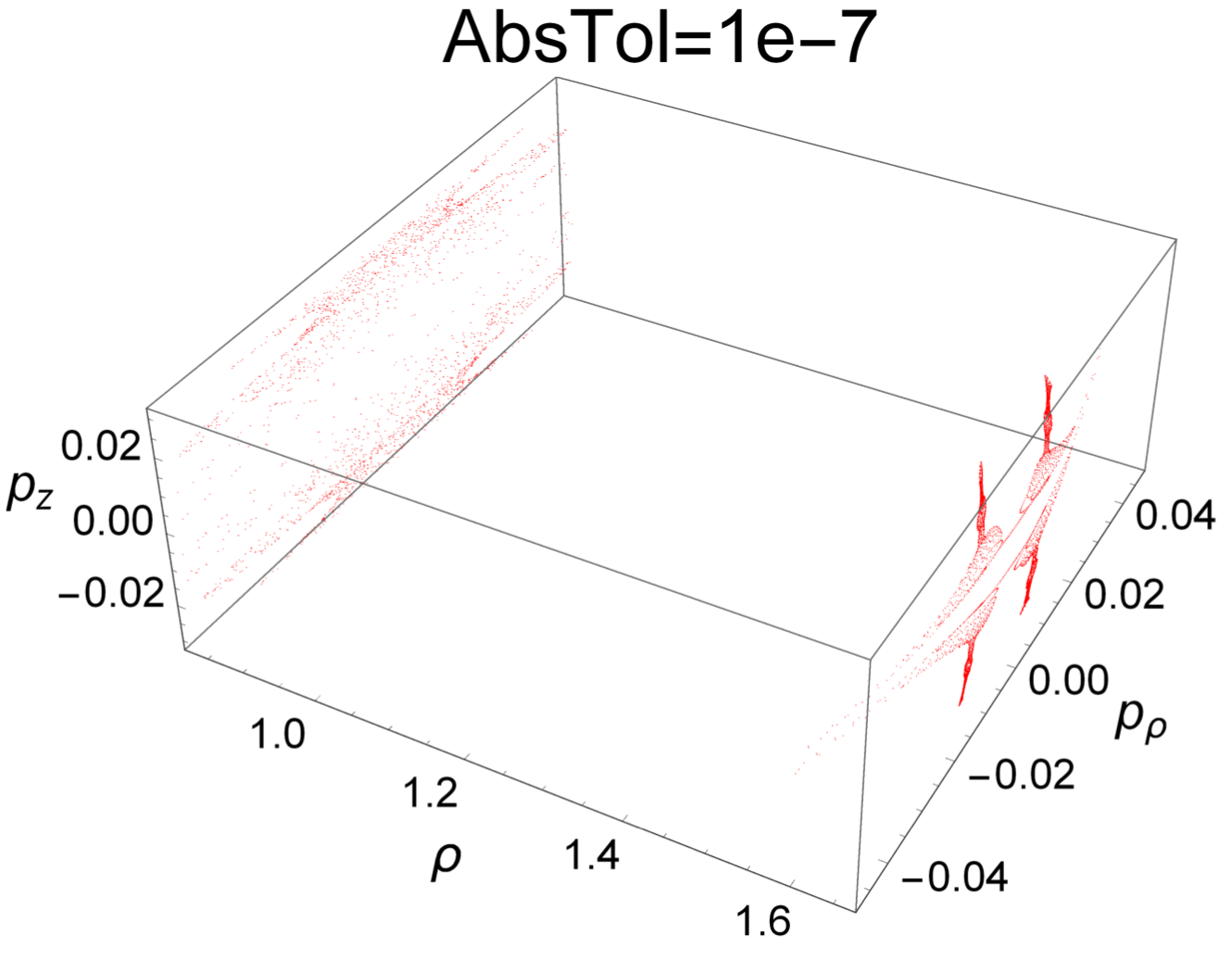}
\includegraphics[width=0.43\linewidth, angle=0.0]{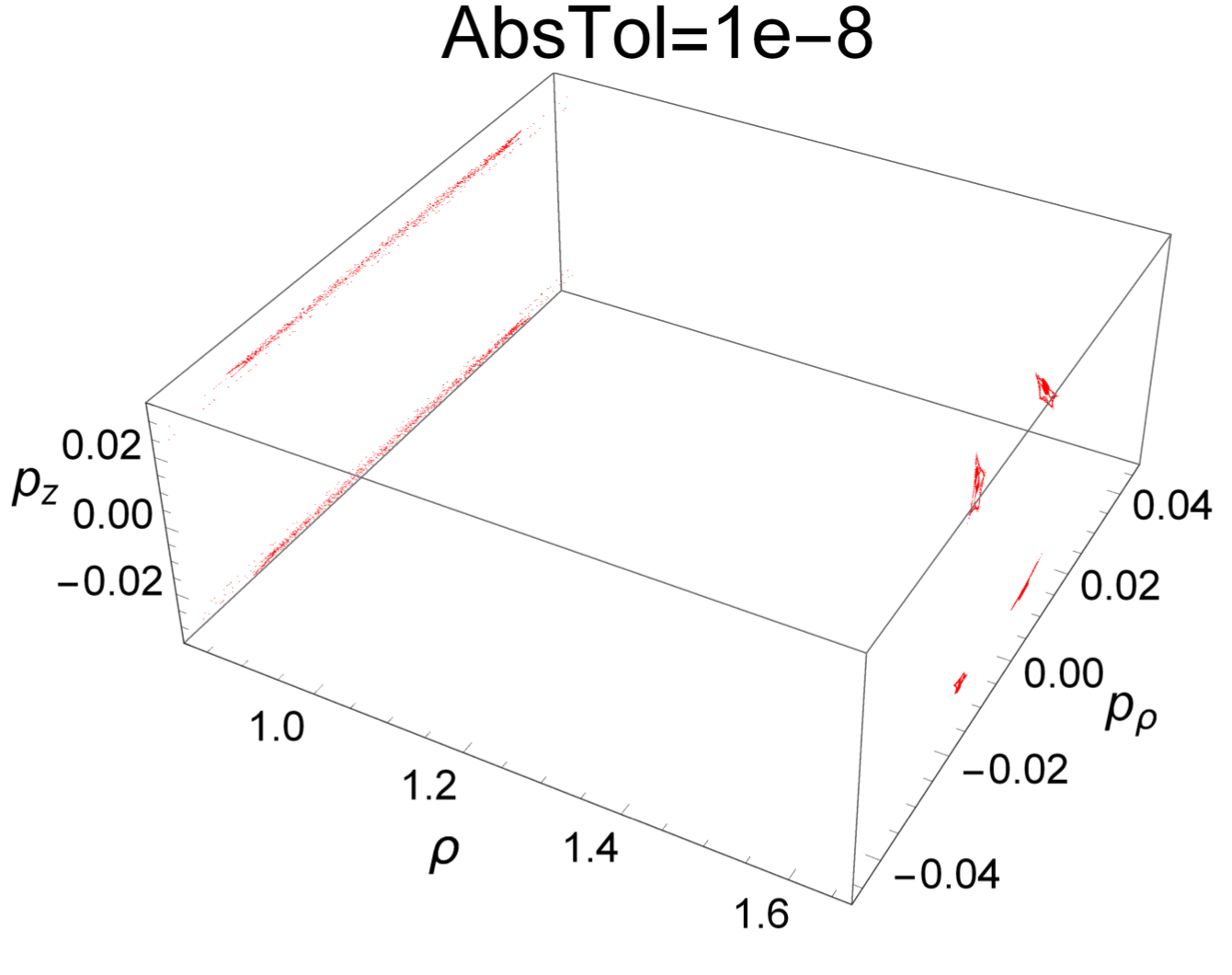}\\
\includegraphics[width=0.43\linewidth, angle=0.0]{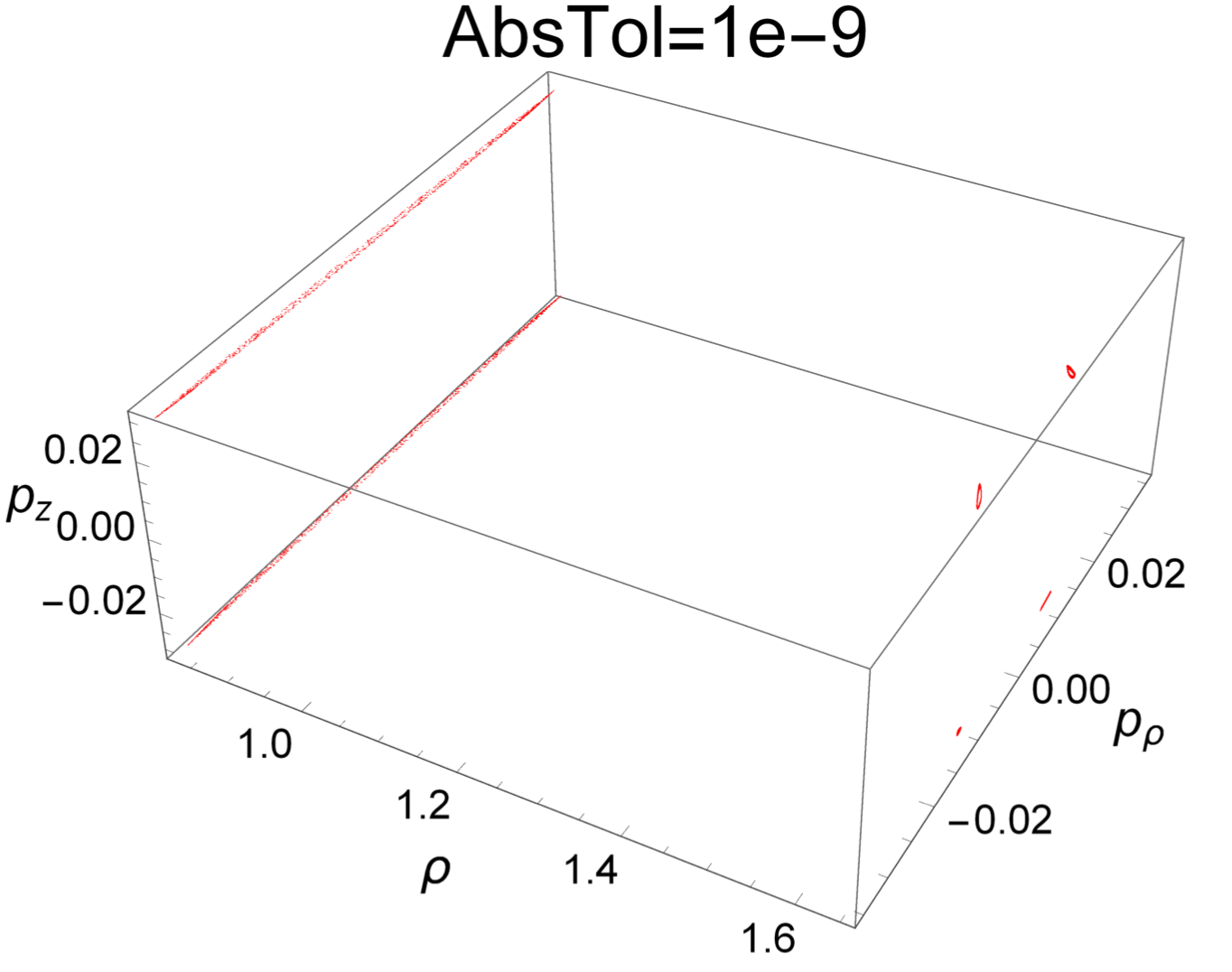}
\includegraphics[width=0.43\linewidth,angle=0.0]{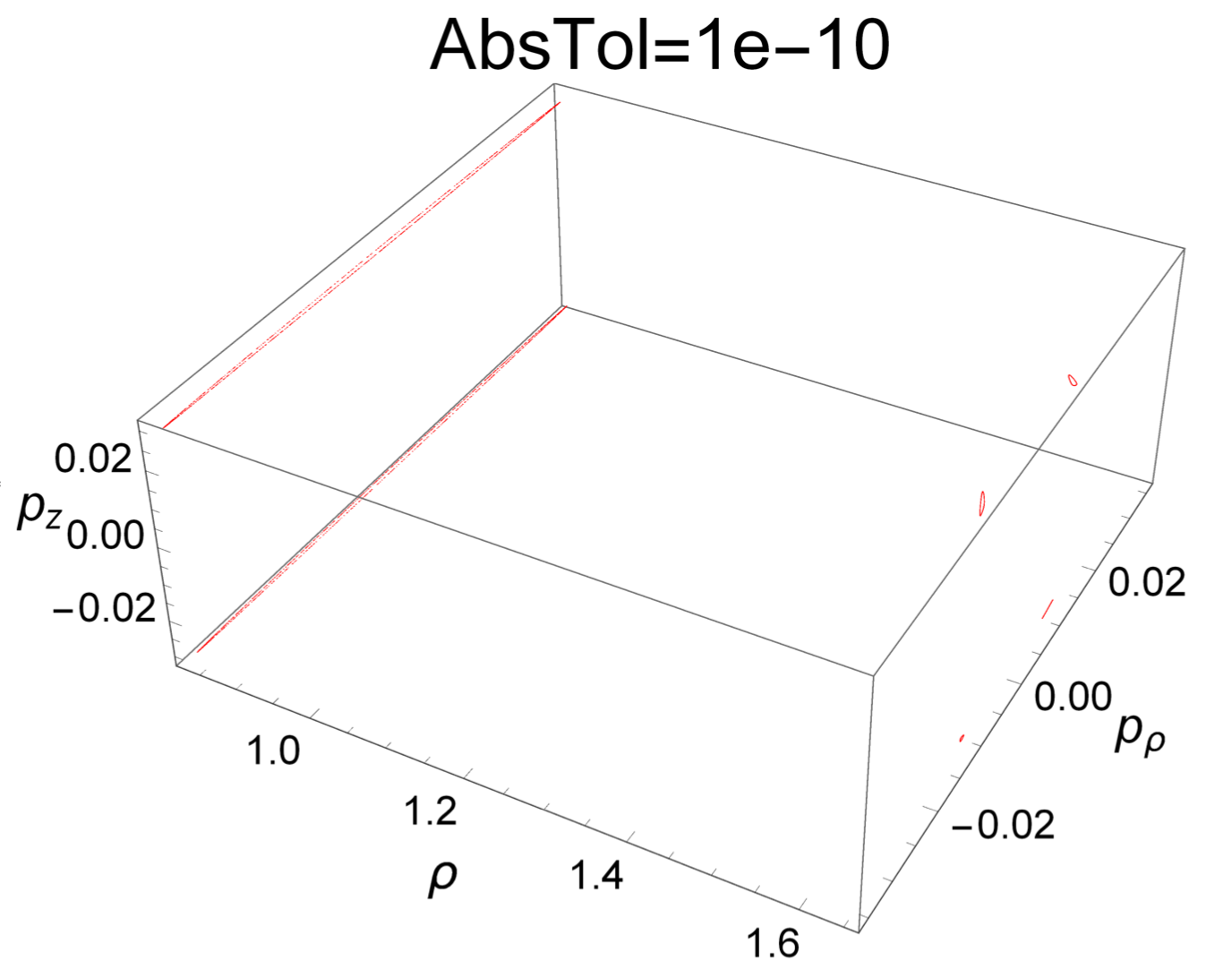}

\caption{\label{figa2} Dependence of the 3D Poincar\'{e} maps of the case shown in Figure \ref{fig6} on the absolute tolerance of the numerical calculations. The relative tolerance is $10^{-10}$.}
\end{figure*}

{\bf 
To estimate the numerical error of our calculations, we plot the 3D Poincar\'{e} maps.  With the initial conditions the same as Figure \ref{fig6}, i.e., $z=0.03829$, $\rho= 0.8307$, $p_z=0$, $p_\rho = 0$, and a relative tolerance of  $10^{-10}$, Figure \ref{figa2} shows the dependence of the 3D Poincar\'{e} map on the absolute tolerance. We found that the result converges to 3 pairs of loops when the relative tolerance is less than $10^{-8}$ and the absolute tolerance is less than $10^{-9}$. Then their projection onto the \{$\rho,\ p_\rho$\} plane should also be loops for the energy conservation. By adjusting the initial conditions, we actually found the periodic orbit associated with this quasi-periodic orbit. Therefore segments of the 2D Poincar\'{e} maps presented in the text are just squeezed loops around a fixed point of the associated stable periodic orbit.

}

\section*{references}
\bibliography{sorsamp} 

\end{document}